\documentclass[aps,prd,twocolumn,superscriptaddress]{revtex4}
\usepackage{epsfig,epsf}
\usepackage{amsmath}
\usepackage{amsthm}
\usepackage{amsfonts}
\usepackage{amssymb}
\usepackage{dsfont}
\usepackage{multirow}
\usepackage{appendix}
\usepackage{slashed}
\usepackage[active]{srcltx}
\usepackage{psfrag}

\begin{document}

\title{Heavy exotic scalar meson $T_{bb;\overline{u}\overline{s}}^{-}$ }
\author{S.~S.~Agaev}
\affiliation{Institute for Physical Problems, Baku State University, Az--1148 Baku,
Azerbaijan}
\author{K.~Azizi}
\affiliation{Department of Physics, University of Tehran, North Karegar Avenue, Tehran
14395-547, Iran}
\affiliation{Department of Physics, Do\v{g}u\c{s} University, Acibadem-Kadik\"{o}y, 34722
Istanbul, Turkey}
\author{B.~Barsbay}
\affiliation{Department of Physics, Do\v{g}u\c{s} University, Acibadem-Kadik\"{o}y, 34722
Istanbul, Turkey}
\affiliation{Department of Physics, Kocaeli University, 41380 Izmit, Turkey}
\author{H.~Sundu}
\affiliation{Department of Physics, Kocaeli University, 41380 Izmit, Turkey}

\begin{abstract}
The spectroscopic parameters and decay channels of the scalar tetraquark $%
T_{bb;\overline{u}\overline{s}}^{-}$ (in what follows $T_{b:\overline{s}%
}^{-} $) are investigated in the framework of the QCD sum rule method. The
mass and coupling of the $T_{b:s}^{-}$ are calculated using the two-point
sum rules by taking into account quark, gluon and mixed vacuum condensates
up to dimension 10. Our result for its mass $m=(10250 \pm 270)~\mathrm{MeV}$
demonstrates that $T_{b:\overline{s}}^{-} $ is stable against the strong and
electromagnetic decays. Therefore to find the width and mean lifetime of the
$T_{b:\overline{s}}^{-}$, we explore its dominant weak decays generated by
the transition $b \to W^{-}c$. These channels embrace the semileptonic decay
$T_{b:\overline{s}}^{-} \to Z_{bc;\overline{u}\overline{s}}^{0}l\overline{%
\nu }_{l}$ and nonleptonic modes $T_{b:\overline{s}}^{-} \to Z_{bc;\overline{%
u}\overline{s}}^{0}\pi^{-}(K^{-}, D^{-}, D_s^{-})$, which at the final state
contain the scalar tetraquark $Z_{bc;\overline{u}\overline{s}}^{0}$. Key
quantities to compute partial widths of the weak decays are the form factors
$G_1(q^2)$ and $G_2(q^2)$: they determine differential rate $d\Gamma/dq^2$
of the semileptonic and partial widths of the nonleptonic processes,
respectively. These form factors are extracted from relevant three-point sum
rules at momentum transfers $q^2$ accessible for such analysis. By means of
the fit functions $F_{1(2)}(q^2)$ they are extrapolated to cover the whole
integration region $m_l^{2}\leq q2\leq(m-\widetilde m)^2$, where $\widetilde
m$ is the mass of $Z_{bc;\overline{u}\overline{s}}^{0}$. Predictions for the
full width $\Gamma _{\mathrm{full}}=(15.21\pm 2.59)\times 10^{-10}~\mathrm{%
MeV}$ and mean lifetime $4.33_{-0.63}^{+0.89}\times 10^{-13}~\mathrm{s}$ of
the $T_{b:s}^{-} $ are useful for experimental and theoretical
investigations of this exotic meson.
\end{abstract}

\maketitle


\section{Introduction}

\label{sec:Int}
Investigation of exotic mesons that are composed of four quarks
(tetraquarks) is among  the interesting topics of the high energy physics.
Experimental information collected during last years by various
collaborations and theoretical progress achieved in the framework of
different methods and models form rapidly growing field of exotic studies
\cite%
{Chen:2016qju,Chen:2016spr,Esposito:2016noz,Olsen:2017bmm,Brambilla:2019esw}.

The states observed in experiments till now and interpreted as candidates to
exotic mesons have different natures. Thus, some of them are neutral
charmonium (bottomonium)-like resonances and may be considered as excited
states of the charmonium. Others bear an electric charge and are free of
these problems, but reside close to two-meson thresholds permitting an
interpretation as bound states of conventional mesons or dynamical effects.
It is worth noting that all of the discovered tetraquarks have large full
widths and decay strongly to two conventional mesons. Therefore, four-quark
compounds stable against strong and electromagnetic interactions, and
decaying only through weak transformations can provide valuable information
on tetraquarks.

The stability of the tetraquarks $QQ^{\prime }\overline{q}\overline{q}%
^{\prime }$ (in what follows denoted as $T_{QQ^{\prime };\overline{q}%
\overline{q}^{\prime }}$) were studied already in original articles \cite%
{Ader:1981db,Lipkin:1986dw,Zouzou:1986qh,Carlson:1987hh}, in which it was
proved that a heavy $Q^{(\prime )}$ and light $q^{(\prime )}$ quarks may
form the stable exotic mesons provided the ratio $m_{Q}/m_{q}$ is large
enough. In fact, the isoscalar axial-vector tetraquark $\ T_{bb;\overline{u}%
\overline{d}}^{-}$ with the mass lower than the $B\overline{B}^{\ast }$
threshold is a strong-interaction stable state \cite{Carlson:1987hh}.

These problems were addressed in numerous later publications using for
investigations various approaches, including the chiral, the dynamical, and
the relativistic quark models. Computational tools employed in these
investigations encompassed all diversity of methods available in the high
energy physics. Thus, quark models were used in Refs.\ \cite%
{Pepin:1996id,Janc:2004qn,Cui:2006mp,Vijande:2006jf,Ebert:2007rn} to explore
features and calculate parameters of the states $T_{QQ}$. These tetraquarks
were analyzed in the framework of the QCD two-point sum rule method, as well
\cite{Navarra:2007yw,Dias:2011mi}. The masses of the axial-vector states $%
T_{bb;\overline{u}\overline{d}}^{-}$ and $T_{cc;\overline{u}\overline{d}%
}^{+} $ were extracted in Ref.\ \cite{Navarra:2007yw}. In accordance with
results of this work, the mass of the tetraquark $T_{bb;\overline{u}%
\overline{d}}^{-} $ amounts to $10.2\pm 0.3\ \mathrm{GeV}$, which is below
the open bottom threshold. In other words, this particle is stable against
strong decays. Parameters of the states $QQ\overline{q}\overline{q}$ with
the spin-parity $J^{\mathrm{P}}=0^{-},\ 0^{+},\ 1^{-}$ and $1^{+}$ were
found in the framework of the sum rule method in Ref.\ \cite{Du:2012wp}.
There are publications in the literature devoted to investigation of
production mechanisms of the tetraquarks $T_{cc}$ in the heavy ion and
proton-proton collisions, in electron-positron annihilations, in $B_{c}$
meson and heavy $\Xi _{bc}$ baryon decays, and to analysis of their possible
decay channels \cite%
{SchaffnerBielich:1998ci,DelFabbro:2004ta,Lee:2007tn,Hyodo:2012pm,Esposito:2013fma}%
.

The discovery of the doubly charmed baryon $\Xi _{cc}^{++}=ccu$ by the LHCb
Collaboration \cite{Aaij:2017ueg} generated new studies of double-heavy
tetraquarks \cite%
{Karliner:2017qjm,Luo:2017eub,Eichten:2017ffp,Wang:2017dtg,Ali:2018ifm,
Ali:2018xfq,Junnarkar:2018twb,Tang:2019nwv}. Investigations prove that
double-charm exotic mesons are unstable against the strong and
electromagnetic decays. Thus, in Ref.\ \cite{Karliner:2017qjm} it was shown
that, the mass of the axial-vector tetraquark $T_{cc\overline{u}\overline{d}%
}^{+}$ is equal to $(3882\pm 12)\ \mathrm{MeV}$, which is above thresholds
for decays to $D^{0}D^{\ast +}$ and $D^{0}D^{+}\gamma $ final states. The
states $T_{cc;\overline{s}\overline{s}}^{++}$ and $T_{cc;\overline{d}%
\overline{s}}^{++}$ that belong to the class of doubly charged tetraquarks
were investigated in our article \cite{Agaev:2018vag}. Performed analysis
demonstrated that, masses of these four-quark compounds are above the $%
D_{s}^{+}D_{s0}^{\ast +}(2317)$ and $D^{+}D_{s0}^{\ast +}(2317)$ thresholds,
and they can decay to these conventional mesons. The widths of these strong
decays, evaluated also in Ref.\ \cite{Agaev:2018vag}, allowed us to classify
the tetraquarks $T_{cc;\overline{s}\overline{s}}^{++}$ and $T_{cc;\overline{d%
}\overline{s}}^{++}$ as relatively broad resonances.

The double-beauty tetraquarks $bb\overline{q}\overline{q}^{\prime }$ are
particles of special interest, because in this case the ratio $%
m_{b}/m_{q(q^{\prime })}$ reaches its maximum value, and they may form
stable compositions. Indeed, the mass of the axial-vector state $T_{bb;%
\overline{u}\overline{d}}^{-}$ was reevaluated in Ref.\ \cite%
{Karliner:2017qjm} using a phenomenological model and experimental
information of the LHCb collaboration \cite{Aaij:2017ueg}. In accordance
with results of this work the mass of the isoscalar axial-vector state $%
T_{bb;\overline{u}\overline{d}}^{-}$ equals to $m=(10389\pm 12)~\mathrm{MeV}$
which is $215~\mathrm{MeV}$ below the $B^{-}\overline{B}^{\ast 0}$ threshold
and $170~\mathrm{MeV}$ below the threshold for decay $B^{-}\overline{B}%
^{0}\gamma $. This means that the tetraquark $T_{bb;\overline{u}\overline{d}%
}^{-}$ is stable against the strong and electromagnetic decays and
transforms to ordinary mesons only through weak processes. The conclusion
about the strong-interaction stability of the tetraquarks $T_{bb;\overline{u}%
\overline{d}}^{-}$, $T_{bb;\overline{u}\overline{s}}^{-}$, and $T_{bb;%
\overline{d}\overline{s}}^{0}$ was made in Ref.\ \cite{Eichten:2017ffp} on
the basis of the relations extracted from heavy-quark symmetry.The mass $%
m=10482~\mathrm{MeV}$ of the axial-vector tetraquark $T_{bb;\overline{u}%
\overline{d}}^{-}$ found there is $121~\mathrm{MeV}$ below the open-bottom
threshold.

In Ref.\ \cite{Agaev:2018khe} we computed the spectroscopic parameters of
the axial-vector tetraquark $T_{bb;\overline{u}\overline{d}}^{-}$ by means
of the QCD sum rule method. Our result for the mass of this particle $%
m=(10035~\pm 260)~\mathrm{MeV}$ confirmed once more that it is stable
against the strong and electromagnetic decays. In this paper, we evaluated
also the total width and mean lifetime of $T_{bb;\overline{u}\overline{d}%
}^{-}$ using its semileptonic decay channels (see, also Ref.\ \cite%
{Hernandez:2019eox}). The predictions $\Gamma =(7.17\pm 1.23)\times 10^{-8}~%
\mathrm{MeV}$ and $\tau =9.18_{-1.34}^{+1.90}~\mathrm{fs}$ provide
information useful for experimental investigation of the double-beauty
exotic mesons.

The axial-vector four-quark systems $qq^{\prime }\overline{Q}\overline{Q}$,
where $Q$ is one of the heavy $b$ or $c$ quarks and $q$, $q^{\prime }$ are
light quarks were explored in Ref.\ \cite{Tang:2019nwv}. In this work, the
authors considered the octet-octet $[\mathbf{8}_{c}]_{q\overline{Q}}\otimes
\lbrack \mathbf{8}_{c}]_{q^{\prime }\overline{Q}}$ and singlet-singlet $[%
\mathbf{1}_{c}]_{q\overline{Q}}\otimes \lbrack \mathbf{1}_{c}]_{q^{\prime }%
\overline{Q}}$ color configurations and calculated masses of these
tetraquarks by means of the QCD sum rule method. Obtained predictions for
the masses of the octet-octet tetraquarks $ud\overline{b}\overline{b}$ and $%
us\overline{b}\overline{b}$ are above corresponding two-meson thresholds,
and hence these states can decay through strong interactions. The molecular
or color singlet-singlet tetraquarks with masses $(10360~\pm 150)~\mathrm{MeV%
}$ for $ud\overline{b}\overline{b}$ and $(10480~\pm 150)~\mathrm{MeV}$ for $%
us\overline{b}\overline{b}$ seem are stable particles.

It turned out that not only exotic mesons containing $bb$ diquarks, but also
tetraquarks built of $bc$ may be stable against the strong and
electromagnetic decays (see, Refs.\ \cite%
{Karliner:2017qjm,Eichten:2017ffp,Agaev:2018khe,Francis:2018jyb,Sundu:2019feu}%
). Thus, analysis of Ref.\ \cite{Agaev:2018khe} proved that the scalar
tetraquark $Z_{bc;\overline{u}\overline{d}}^{0}$ has the mass $%
m_{Z}=(6660\pm 150)~\mathrm{MeV}$, which is considerably below thresholds
for strong and electromagnetic decays. In other words, $Z_{bc;\overline{u}%
\overline{d}}^{0}$ transforms due to weak decays that allowed us to estimate
in Ref.\ \cite{Sundu:2019feu} its full width and mean lifetime. A situation
with the axial-vector tetraquark $T_{bc;\overline{u}\overline{d}}^{0}$
remains unclear: the mass of this state predicted in the range $(1705\pm
155)~\mathrm{MeV}$ admits twofold explanations \cite{Agaev:2019kkz}. Indeed,
using the central value of the mass one see that it lies below thresholds
for the strong and electromagnetic decays, whereas the maximum estimate for
the mass $7260$ $\mathrm{MeV}$ is higher than thresholds for strong and
electromagnetic decays to $B^{\ast -}D^{+}/\overline{B}^{\ast 0}D^{0}$ and $%
D^{+}B^{-}\gamma /D^{0}\overline{B}^{0}\gamma $, respectively. In the first
case the width and lifetime of the tetraquark $T_{bc;\overline{u}\overline{d}%
}^{0}$ are determined by its weak decays. In the second scenario the width
of $T_{bc;\overline{u}\overline{d}}^{0}$ is fixed mainly by strong modes,
because widths of weak and electromagnetic processes are small and can be
ignored \cite{Agaev:2019kkz}.

It is worth noting that some of heavy exotic mesons containing diquarks $bs$
may be stable as well. Thus, the scalar tetraquark $T_{bs;\overline{u}%
\overline{d}}^{-}$ is strong- and electromagnetic-interaction stable
particle: its spectroscopic parameters and semileptonic decays were explored
in Ref.\ \cite{Agaev:2019wkk}.

In the present article we study the scalar tetraquark $T_{b:\overline{s}%
}^{-} $ with the quark content $bb\overline{u}\overline{s}$ and compute its
spectroscopic parameters, full width and mean lifetime. The mass $m$ and
coupling $f$ of $T_{b:\overline{s}}^{-}$ are extracted from the QCD
two-point sum rules by taking into account vacuum expectation values of the
local quark, gluon and mixed operators up to dimension ten. The information
on the mass of this state is crucial to determine whether $T_{b:\overline{s}%
}^{-}$ is strong- and electromagnetic-interaction stable particle or not. It
is not difficult to see that dissociation to a pair of conventional
pseudoscalar mesons $B^{-}\overline{B}_{s}^{0}$ is the first $S$-wave strong
decay channel for the unstable $T_{b:\overline{s}}^{-}$. Therefore if the
mass of $T_{b:\overline{s}}^{-}$ is higher than the $B^{-}\overline{B}%
_{s}^{0}$ threshold $10646~\mathrm{MeV}$ then one should calculate the width
of the process $T_{b:\overline{s}}^{-}\rightarrow B^{-}\overline{B}_{s}^{0}$%
. But, our investigations show (see, below) that the mass of the tetraquark $%
T_{b:\overline{s}}^{-}$ is equal to $m=(10250\pm 270)~\mathrm{MeV,}$ and
lies below this bound. The $T_{b:\overline{s}}^{-}$ is stable against the
possible electromagnetic transition $T_{b:\overline{s}}^{-}\rightarrow B^{-}%
\overline{B}_{s1}(5830)\gamma $ as well, because for realization of this
process the mass of the master particle should exceed $11108\ \mathrm{MeV}$
which is not a case. Therefore to evaluate the full width and lifetime of $%
T_{b:\overline{s}}^{-}$ one has to explore its weak decays.

The weak transformations of the $T_{b:\overline{s}}^{-}$ may run due to the
subprocesses $b\rightarrow W^{-}c$, and $b\rightarrow W^{-}u$ which generate
its semileptonic dissociation to scalar four-quark mesons $Z_{bc;\overline{u}%
\overline{s}}^{0}$ (hereafter $Z_{b:\overline{s}}^{0}$) and $Z_{bu;\overline{%
u}\overline{s}}^{0}$. The process $T_{b:\overline{s}}^{-}\rightarrow Z_{b:%
\overline{s}}^{0}l\overline{\nu }_{l}$ is dominant weak channel for $T_{b:%
\overline{s}}^{-}$, because the decay $T_{b:\overline{s}}^{-}\rightarrow
Z_{bu;\overline{u}\overline{s}}^{0}l\overline{\nu }_{l}$ is suppressed
relative to first one by a factor $|V_{ub}|^{2}/|V_{cb}|^{2}\approx 0.01$
with $|V_{q_{1}q_{2}}|$ being the Cabibbo-Kobayasi-Maskawa (CKM) matrix
elements.

But the subprocess $b\rightarrow W^{-}c$ can also give rise to nonleptonic
weak decays of the $T_{b:\overline{s}}^{-}$. Indeed, the vector boson $W^{-}$
instead of a lepton pair $l\overline{\nu }_{l}$ can produce $d\overline{u},\
s\overline{u},\ d\overline{c}$, and $s\overline{c}$ quarks as well. These
quarks afterwards form one of the conventional mesons $M=\pi ^{-}$, $K^{-}$,
$D^{-}$ and $D_{s}^{-}$ leading to the nonleptonic final states $Z_{b:%
\overline{s}}^{0}M$. Depending on the difference $m-\widetilde{m}$, where $%
\widetilde{m}$ is the $Z_{b:\overline{s}}^{0}$ tetraquark's mass, some or
all of these nonleptonic weak decays become kinematically allowed.

We calculate the full width of the $T_{b:\overline{s}}^{-}$ by taking into
account its semileptonic and nonleptonic decay modes. To this end, employing
the QCD three-point sum rule approach, we determine the weak form factors $%
G_{1(2)}(q^{2})$ necessary to evaluate the differential rates of the
semileptonic decays. Partial width of the processes $T_{b:\overline{s}%
}^{-}\rightarrow Z_{b:\overline{s}}^{0}l\overline{\nu }_{l}$, $l=e^{-}$, $%
\mu ^{-}$ and $\tau ^{-}$ can be found by integrating the differential rates
over kinematically allowed momentum transfers $q^{2}$, whereas width of the
nonleptonic decays $T_{b:\overline{s}}^{-}\rightarrow Z_{b:\overline{s}%
}^{0}M $ are fixed by values of the $G_{1(2)}($ $q^{2})$ at $q^{2}=m_{M}^{2}$%
, where $m_{M}$ is the mass of a produced meson.

This article is structured in the following way: In Section \ref{sec:Masses}%
, we calculate the spectroscopic parameters of the scalar tetraquarks $T_{b:%
\overline{s}}^{-}$ and $Z_{b:\overline{s}}^{0}$. For these purposes, we
derive two-point sum rules from analysis of corresponding correlation
functions and include into calculations the quark, gluon and mixed
condensates up to dimension ten. In Section \ref{sec:Decays1}, we derive
three-point sum rules for the weak form factors $G_{1(2)}(q^{2})$ and
compute them in regions of the momentum transfer, where the method gives
reliable predictions. We extrapolate $G_{1(2)}(q^{2})$ to the whole
integration region by means of fit functions and find partial widths of the
semileptonic decays $T_{b:\overline{s}}^{-}\rightarrow Z_{b:\overline{s}%
}^{0}l\overline{\nu }_{l}$,\ where $l=e^{-}$, $\mu ^{-}$ and $\tau ^{-}$. In
Section \ref{sec:Decays2} we analyze the nonleptonic weak decays $T_{b:%
\overline{s}}^{-}\rightarrow Z_{b:\overline{s}}^{0}M$ of the tetraquark $%
T_{b:\overline{s}}$. Here we also present our final estimate for the full
width and mean lifetime of the $T_{b:\overline{s}}$. Section \ref{sec:Disc}
is reserved for discussion and concluding notes. Appendix contains explicit
expressions of quark propagators, and the correlation function used to
evaluate parameters of the tetraquark $T_{b:\overline{s}}^{-}$.


\section{Mass and coupling of the scalar tetraquarks $T_{b:\overline{s}}^{-}$
and $Z_{b:\overline{s}}^{0}$}

\label{sec:Masses}
The spectroscopic parameters of the tetraquark $T_{b:\overline{s}}^{-}$ are
necessary to reveal its nature and answer questions about its stability. The
mass and coupling of $Z_{b:\overline{s}}^{0}$ are important to explore the
weak decays of the master particle $T_{b:\overline{s}}^{-}$. It is worth
noting that the $T_{b:\overline{s}}^{-}$ and $Z_{b:\overline{s}}^{0}$ have
the same heavy diquark-light antidiquark organization.

The parameters of these states can be extracted from the two-point
correlation function
\begin{equation}
\Pi (p)=i\int d^{4}xe^{ipx}\langle 0|\mathcal{T}\{J(x)J^{\dag
}(0)\}|0\rangle,  \label{eq:CF1}
\end{equation}%
where $J(x)$ is the interpolating current for a scalar particle. It is known
that interpolating currents for hadrons, including exotic mesons, should be
colorless constructions. A diquark $QQ^{\prime }$ can belong either to
color-antitriplet $[\overline{\mathbf{3}}_{c}]_{QQ^{\prime }}$ or
color-sextet $[\mathbf{6}_{c}]_{QQ^{\prime }}$ representation of the group $%
SU_{c}(3)$. Accordingly, an antidiquark $\overline{q}\overline{q}^{\prime }$
has triplet or antisextet color structures. Colorless currents of exotic
mesons should have color organizations $[\overline{\mathbf{3}}%
_{c}]_{QQ^{\prime }}\otimes \lbrack \mathbf{3}_{c}]_{\overline{q}\overline{q}%
^{\prime }}$ or $[\mathbf{6}_{c}]_{QQ^{\prime }}\otimes \lbrack \overline{%
\mathbf{6}}_{c}]_{\overline{q}\overline{q}^{\prime }}$. The color and flavor
antisymmetric scalar diquarks are most attractive and stable two-quark
structures \cite{Jaffe:2004ph}. But, because the heavy diquark $bb$ in $T_{b:%
\overline{s}}^{-}$ contains two quarks of the same flavor, relevant diquark
field has to be symmetric in color indices, i.e., belong to sextet
representation of the color group. The interpolating current for $T_{b:%
\overline{s}}^{-}$ built of color-sextet scalar diquark and antidiquark
fields has the following form \cite{Du:2012wp}
\begin{equation}
J(x)=[b_{a}^{T}(x)C\gamma _{5}b_{b}(x)][\overline{u}_{a}(x)\gamma _{5}C%
\overline{s}_{b}^{T}(x)],  \label{eq:CR1}
\end{equation}%
where $a$, and $b$ are color indices and $C$ is the charge-conjugation
operator. The second term $[b_{a}^{T}(x)C\gamma _{5}b_{b}(x)][\overline{u}%
_{b}(x)\gamma _{5}C\overline{s}_{a}^{T}(x)]$ in $J(x)$ is equal to the one
presented in Eq.\ (\ref{eq:CR1}), therefore we use this compact expression
for the interpolating current.

The final-state tetraquark $Z_{b:\overline{s}}^{0}=bc\overline{u}\overline{s}
$ may have color-antisymmetric $[\overline{\mathbf{3}}_{c}]_{bc}\otimes
\lbrack \mathbf{3}_{c}]_{\overline{u}\overline{s}}$ or symmetric $[\mathbf{6}%
_{c}]_{bc}\otimes \lbrack \overline{\mathbf{6}}_{c}]_{\overline{u}\overline{s%
}}$ interpolating currents. Our calculations demonstrate that the tetraquark
$T_{b:\overline{s}}^{-}$ decays weakly only to $Z_{b:\overline{s}}^{0}$ with
color-sextet constituents: a matrix element for weak transition to
color-antisymmetric state $bc\overline{u}\overline{s}$ vanishes identically.
In other words, in weak transitions of $T_{b:\overline{s}}^{-}$ to $Z_{b:%
\overline{s}}^{0}$ color structures of their constituents remain unchanged.
This is true, at least, for tetraquarks under analysis and for currents
employed to interpolate them. Therefore, for $Z_{b:\overline{s}}^{0}$, we
choose also $[\mathbf{6}_{c}]_{bc}\otimes \lbrack \overline{\mathbf{6}}%
_{c}]_{\overline{u}\overline{s}}$ type current using information from Ref.
\cite{Chen:2013aba}
\begin{eqnarray}
\widetilde{J}(x) &=&[b_{a}^{T}(x)C\gamma _{5}c_{b}(x)]\left[ \overline{u}%
_{a}(x)\gamma _{5}C\overline{s}_{b}^{T}(x)\right.  \notag \\
&&+\overline{u}_{b}(x)\gamma _{5}C\overline{s}_{a}^{T}(x)].  \label{eq:CR2}
\end{eqnarray}%
The exotic mesons with internal organizations (\ref{eq:CR1}) and (\ref%
{eq:CR2}) are ground-state particles with color-symmetric diquarks.

Here, we consider in a detailed form computation of the $T_{b:\overline{s}%
}^{-}$ tetraquark's mass $m$ and coupling $f$, and provide final results
for\ $Z_{b:\overline{s}}^{0}$. To derive the sum rules for $m$ and $f$, we
need first to find the phenomenological expression of the correlation
function $\Pi ^{\mathrm{Phys}}(p)$, which should be written down in terms of
the spectroscopic parameters of $T_{b:\overline{s}}^{-}$. Since $T_{b:%
\overline{s}}^{-}$ is a ground-state particle, we use the "ground-state +
continuum" scheme. Then separating contribution of the tetraquark $T_{b:%
\overline{s}}^{-}$ from effects of the higher resonances and continuum
states, we can write
\begin{equation}
\Pi ^{\mathrm{Phys}}(p)=\frac{\langle 0|J|T_{b:\overline{s}}^{-}(p)\rangle
\langle T_{b:\overline{s}}^{-}(p)|J^{\dagger }|0\rangle }{m^{2}-p^{2}}+\ldots
\label{eq:Phen1}
\end{equation}%
The phenomenological function ~$\Pi ^{\mathrm{Phys}}(p)$ is obtained by
inserting into $\Pi (p)$ a full set of scalar four-quark states and
performing integration over $x$.

Calculation of $\Pi ^{\mathrm{Phys}}(p)$ can be finished by employing the
matrix element
\begin{equation}
\langle 0|J|T_{b:\overline{s}}^{-}(p)\rangle =fm.  \label{eq:ME1}
\end{equation}%
After simple manipulations we get%
\begin{equation}
\Pi ^{\mathrm{Phys}}(p)=\frac{f^{2}m^{2}}{m^{2}-p^{2}}+\ldots
\label{eq:Phen2}
\end{equation}%
The correlation function $\Pi ^{\mathrm{Phys}}(p)$ has a trivial Lorentz
structure which is proportional to $\sim I$. Hence, the only term in Eq.\ (%
\ref{eq:Phen2}) is nothing more than the invariant amplitude $\Pi ^{\mathrm{%
Phys}}(p^{2})$ corresponding to this structure.

Now, we have to fix the second component of the sum rule analysis, and
express $\Pi (p)$ in terms of the quark propagators. To this end, we utilize
the explicit expression of the interpolating current $J(x)$, and contract
relevant heavy and light quark fields to get $\Pi ^{\mathrm{OPE}}(p)$. After
these manipulations, we find%
\begin{eqnarray}
&&\Pi ^{\mathrm{OPE}}(p)=i\int d^{4}xe^{ipx}\mathrm{Tr}\left[ \gamma _{5}%
\widetilde{S}_{s}^{b^{\prime }b}(-x)\gamma _{5}S_{u}^{a^{\prime }a}(-x)%
\right]  \notag \\
&&\times \left\{ \mathrm{Tr}\left[ \gamma _{5}\widetilde{S}_{b}^{aa^{\prime
}}(x)\gamma _{5}S_{b}^{bb^{\prime }}(x)\right] +\mathrm{Tr}\left[ \gamma _{5}%
\widetilde{S}_{b}^{ba^{\prime }}(x)\gamma _{5}S_{b}^{ab^{\prime }}(x)\right]
\right\},  \notag \\
&&  \label{eq:QCD1}
\end{eqnarray}%
where $S_{b}(x)$ and $S_{u(s)}(x)$ are the $b$- and $u(s)$-quark
propagators, respectively. Here we also use the shorthand notation
\begin{equation}
\widetilde{S}_{b(u,s)}(x)=CS_{b(u,s)}^{T}(x)C.  \label{eq:Notation}
\end{equation}%
The propagators of heavy and light quarks used in the present work are
collected in Appendix. The nonperturbative part of these propagators
contains vacuum expectation values of various quark, gluon, and mixed
operators which generate a dependence of $\Pi ^{\mathrm{OPE}}(p)$ on
nonperturbative quantities.

To derive the sum rules, we equate the amplitudes $\Pi ^{\mathrm{Phys}%
}(p^{2})$ and $\Pi ^{\mathrm{OPE}}(p^{2})$, and apply to both sides of the
obtained equality the Borel transformation. This operation is necessary to
suppress contributions of higher resonances and continuum states.
Afterwards, we carry out the continuum subtraction using the assumption on
the quark-hadron duality. The expression found by this way, and an equality\
obtained by applying the operator $d/d(-1/M^{2})$ to the first one form a
system which is enough to obtain the sum rules for $m$%
\begin{equation}
m^{2}=\frac{\Pi (M^{2},s_{0})}{\Pi ^{\prime }(M^{2},s_{0})},  \label{eq:Mass}
\end{equation}%
and $f$%
\begin{equation}
f^{2}=\frac{e^{m^{2}/M^{2}}}{m^{2}}\Pi (M^{2},s_{0}).  \label{eq:Coupl}
\end{equation}%
In Eqs.\ (\ref{eq:Mass}) and (\ref{eq:Coupl}) $\Pi (M^{2},s_{0})$ is the
Borel-transformed and subtracted invariant amplitude $\Pi ^{\mathrm{OPE}%
}(p^{2})$, and $\Pi ^{\prime }(M^{2},s_{0})$ equals to
\begin{equation}
\Pi ^{\prime }(M^{2},s_{0})=\frac{d}{d(-1/M^{2})}\Pi (M^{2},s_{0}).
\end{equation}%
In the case under discussion $\Pi (M^{2},s_{0})$ has the following form%
\begin{equation}
\Pi (M^{2},s_{0})=\int_{\mathcal{M}^{2}}^{s_{0}}ds\rho ^{\mathrm{OPE}%
}(s)e^{-s/M^{2}}+\Pi (M^{2}),  \label{eq:InvAmp}
\end{equation}%
where $\mathcal{M}=2m_{b}+m_{s}$. The $\rho ^{\mathrm{OPE}}(s)$ is the
two-point spectral density, whereas second component of the invariant
amplitude $\Pi (M^{2})$ includes nonperturbative contributions calculated
directly from $\Pi ^{\mathrm{OPE}}(p)$. Explicit expression of $\Pi
(M^{2},s_{0})$ is presented in Appendix.

The sum rules for $m$ and $f$ depend on the Borel and threshold parameters $%
M^{2}$ and $s_{0}$, which appear after the Borel transformation and
continuum subtraction procedures, respectively. Both of $M^{2}$ and $s_{0}$
are the auxiliary parameters a proper choice of which depends on the problem
under analysis, and is one of the important problems in the sum rule
computations.
\begin{widetext}

\begin{figure}[h!]
\begin{center}
\includegraphics[totalheight=6cm,width=8cm]{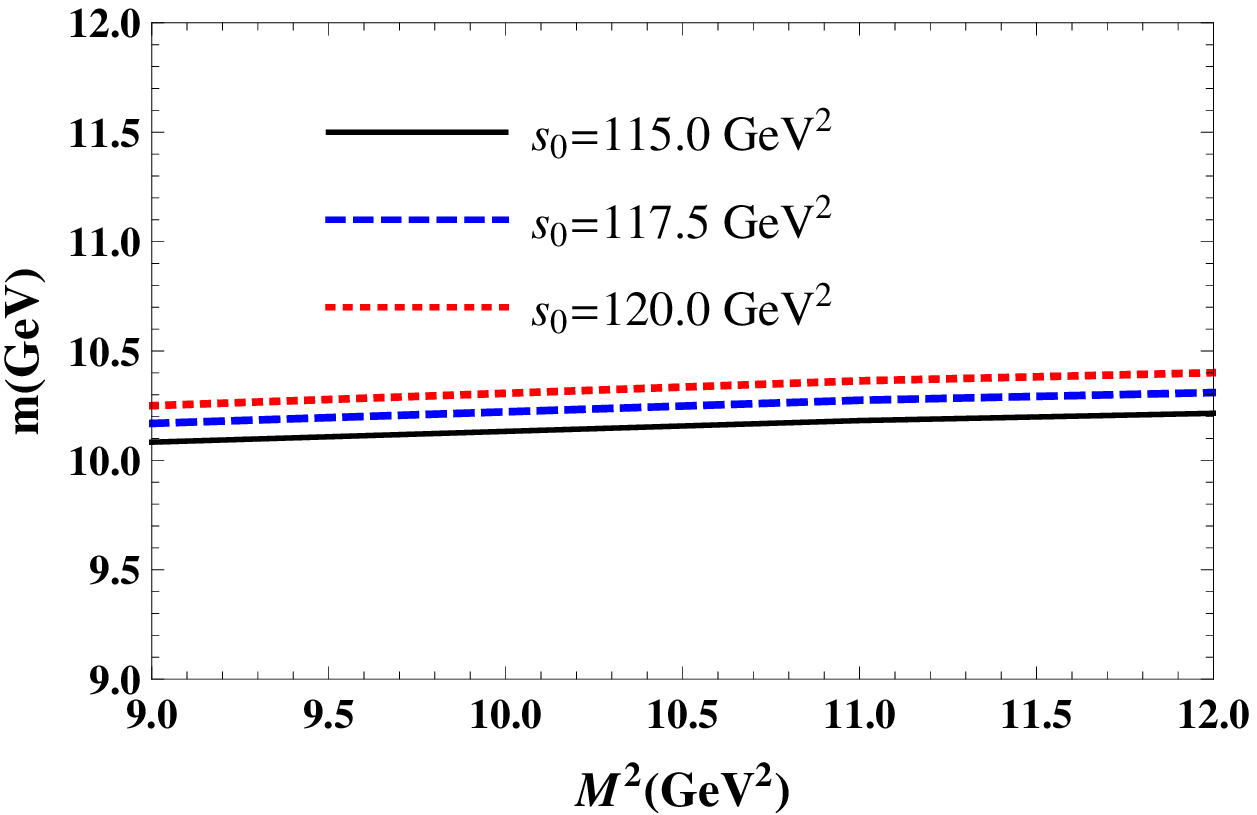}\,\, %
\includegraphics[totalheight=6cm,width=8cm]{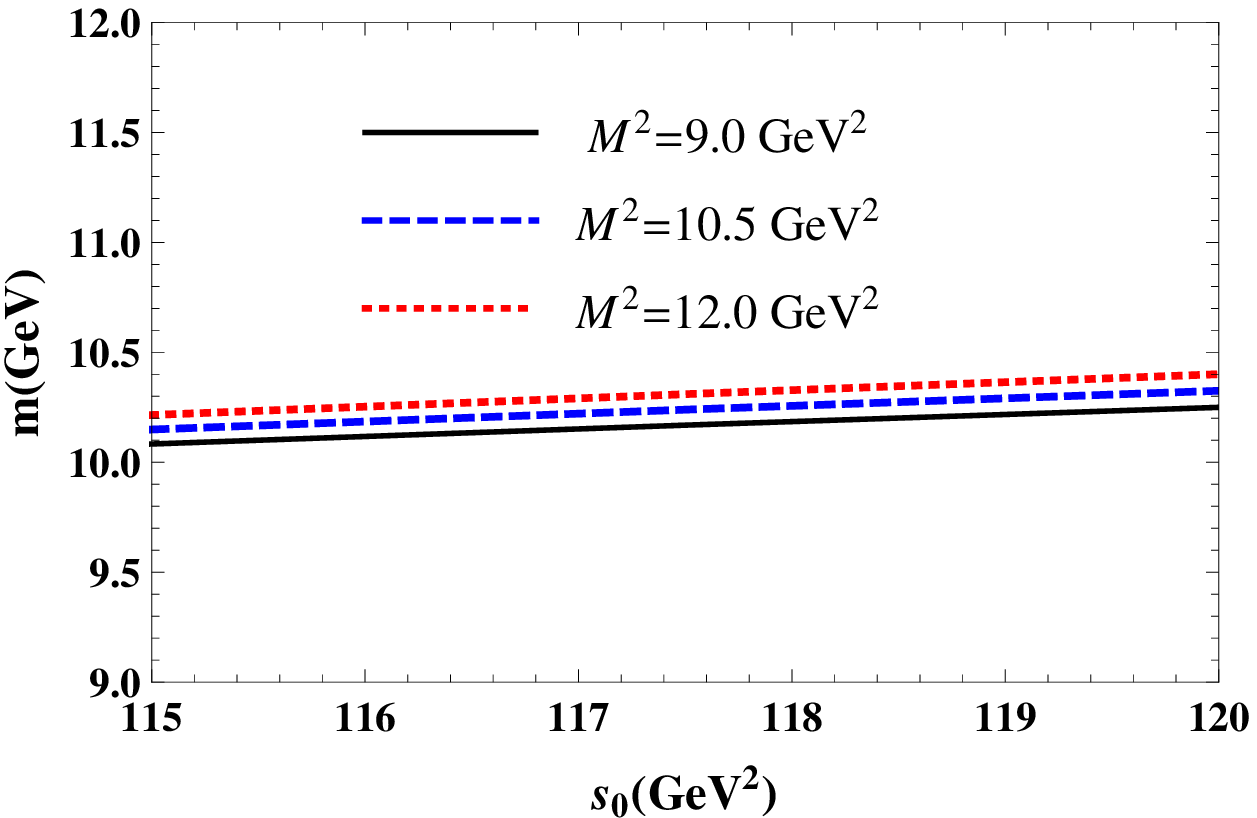}
\end{center}
\caption{ The mass $m$ of the tetraquark $T_{b:\overline{s}}^{-}$ as a
function of the Borel $M^{2}$ (left panel) and continuum threshold $s_{0}$ parameters (right panel).}
\label{fig:MassT}
\end{figure}
\end{widetext}

\begin{widetext}

\begin{figure}[h!]
\begin{center}
\includegraphics[totalheight=6cm,width=8cm]{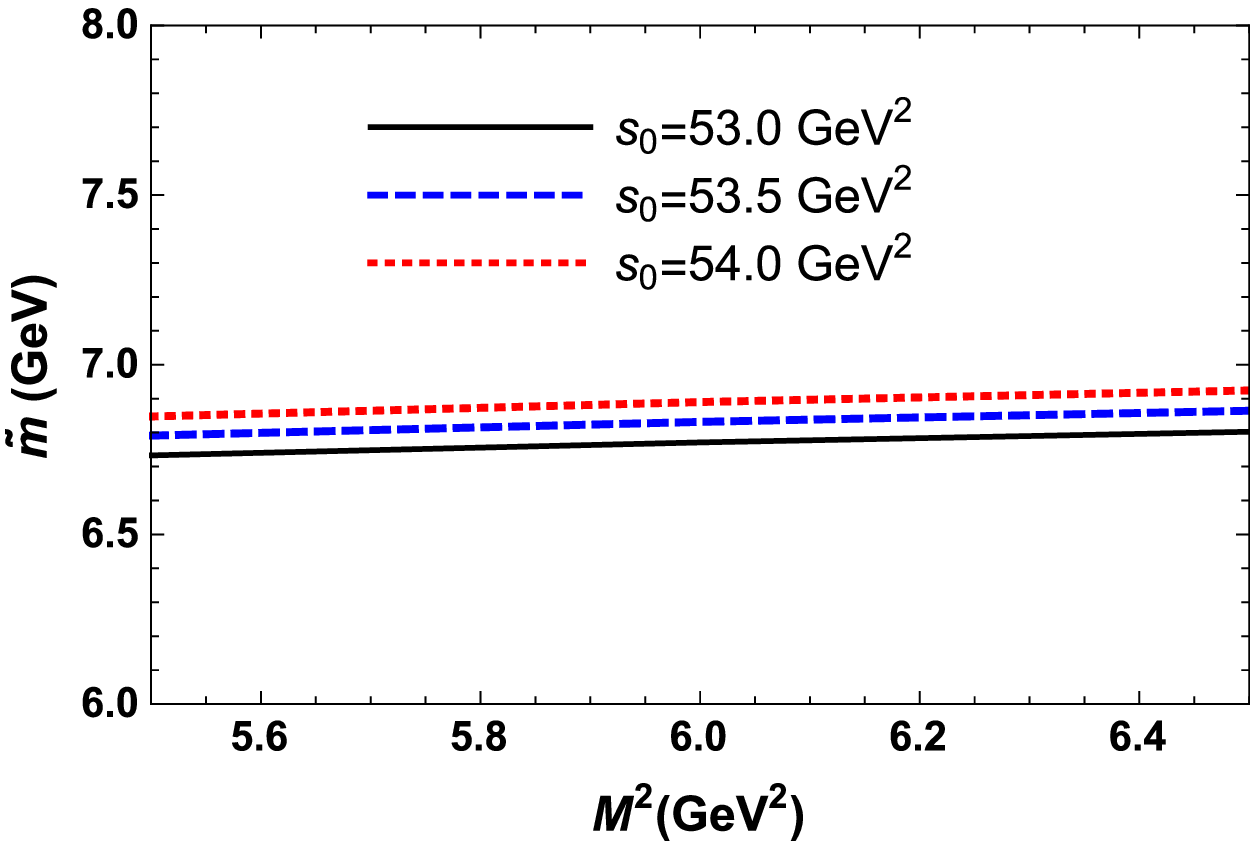}\,\, %
\includegraphics[totalheight=6cm,width=8cm]{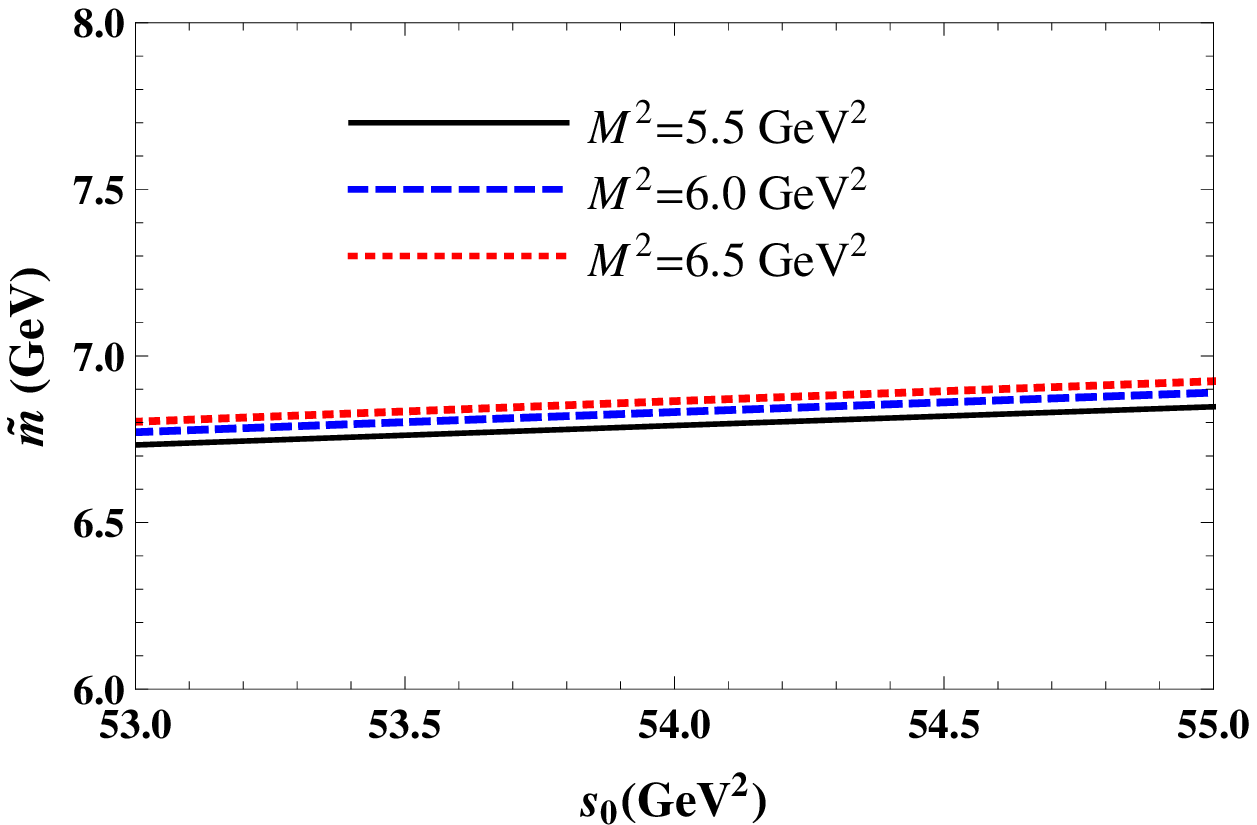}
\end{center}
\caption{ The same as in Fig.\ 1, but for the mass $\widetilde{m}$ of the tetraquark $Z_{b:\overline{s}}^{0}$.}
\label{fig:MassZ}
\end{figure}
\end{widetext}

Apart from $M^{2}$ and $s_{0}$, the sum rules contain also the universal
vacuum condensates and the mass of $b$ and $s$ quarks:%
\begin{eqnarray}
&&\langle \bar{q}q\rangle =-(0.24\pm 0.01)^{3}~\mathrm{GeV}^{3},\ \langle
\bar{s}s\rangle =0.8\langle \bar{q}q\rangle ,  \notag \\
&&\langle \overline{q}g_{s}\sigma Gq\rangle =m_{0}^{2}\langle \overline{q}%
q\rangle ,\ \langle \overline{s}g_{s}\sigma Gs\rangle =m_{0}^{2}\langle \bar{%
s}s\rangle ,  \notag \\
&&m_{0}^{2}=(0.8\pm 0.1)~\mathrm{GeV}^{2},  \notag \\
&&\langle \frac{\alpha _{s}G^{2}}{\pi }\rangle =(0.012\pm 0.004)~\mathrm{GeV}%
^{4},  \notag \\
&&\langle g_{s}^{3}G^{3}\rangle =(0.57\pm 0.29)~\mathrm{GeV}^{6},\
m_{s}=93_{-5}^{+11}~\mathrm{MeV},  \notag \\
&&m_{c}=1.27\pm 0.2~\mathrm{GeV},\ m_{b}=4.18_{-0.02}^{+0.03}~\mathrm{GeV}.
\label{eq:Parameters}
\end{eqnarray}

The working windows for the auxiliary parameters $M^{2}$ and $s_{0}$ have to
satisfy some essential constraints. Thus, at maximum of $M^{2}$ the pole
contribution ($\mathrm{PC}$) should exceed a fixed value, which for the
multiquark systems is chosen in the form
\begin{equation}
\mathrm{PC}=\frac{\Pi (M^{2},s_{0})}{\Pi (M^{2},\infty )}>0.2,  \label{eq:PC}
\end{equation}
The minimum of $M^{2}$ is extracted from analysis of the ratio
\begin{equation}
R(M^{2})=\frac{\Pi ^{\mathrm{DimN}}(M^{2},s_{0})}{\Pi (M^{2},s_{0})}\leq
0.01.  \label{eq:Convergence}
\end{equation}

Fulfilment of Eq.\ (\ref{eq:Convergence}) implies the convergence of the
operator product expansion ($\mathrm{OPE}$) and obtained sum rules. Here, $%
\Pi ^{\mathrm{DimN}}(M^{2},s_{0})$ denotes a contribution to the correlation
function coming from the last term (or a sum of last few terms) in the
expansion. In the present calculations we use a sum of last three terms, and
hence $\mathrm{DimN}$ means $\mathrm{Dim(8+9+10)}$.

The numerical analysis proves that the working regions for the parameters $%
M^{2}$ and $s_{0}$
\begin{equation}
M^{2}\in \lbrack 9,12]\ \mathrm{GeV}^{2},\ s_{0}\in \lbrack 115,120]~\mathrm{GeV}^{2},  \label{eq:Wind1}
\end{equation}%
satisfy all aforementioned constraints on $M^{2}$ and $s_{0}$. Namely, at $%
M^{2}=\ 12\ \mathrm{GeV}^{2}$ the pole contribution is $0.22$, whereas at $%
M^{2}=9~\mathrm{GeV}^{2}$ it amounts to $0.56$. These two values of $M^{2}$
determine the boundaries of a window within of which the Borel parameter can
be varied. At the minimum of $M^{2}=9~\mathrm{GeV}^{2}$ we get $R\approx
0.001$. Apart from that, at the minimum of $M^{2}$ the perturbative
contribution amounts to $85\%$ of the whole result overshooting
significantly the nonperturbative terms.

Our results for $m$ and $f$ are
\begin{eqnarray}
m &=&(10250\pm 270)~\mathrm{MeV},  \notag \\
f &=&(2.69\pm 0.58)\times 10^{-2}~\mathrm{GeV}^{4},  \label{eq:Result1}
\end{eqnarray}%
where we indicate also uncertainties of the computations. These theoretical
errors stem mainly from variation of the parameters $M^{2}$ and $s_{0}$
within allowed limits. It is seen, that for the mass these uncertainties
equal to $\pm 2.6\%$ of its central value, whereas for the coupling $f$ they
are larger and amounts to $\pm 22\%$. In other words, the result for the
mass is less sensitive to the choice of the parameters than the coupling $f$%
. The reason is that the sum rule for the mass (\ref{eq:Mass}) is given as a
ratio of two integrals of the function $\rho ^{\mathrm{OPE}}(s)$ which
stabilizes undesired effects, but even in the situation with the coupling $f$
uncertainties do not exceed limits accepted in sum rule computations. In
Fig.\ \ref{fig:MassT} we plot the sum rule's prediction for $m$ as a
function of the parameters $M^{2}$ and $s_{0}$, where one can see its
residual dependence on them.

The mass and coupling of the scalar tetraquark $Z_{b:\overline{s}}^{0}$ are
calculated by the same way. The QCD side of relevant sum rules is given by
the following formula
\begin{eqnarray}
&&\widetilde{\Pi }^{\mathrm{OPE}}(p)=i\int d^{4}xe^{ipx}\mathrm{Tr}\left[
\gamma _{5}\widetilde{S}_{b}^{aa^{\prime }}(x)\gamma _{5}S_{c}^{bb^{\prime
}}(x)\right]  \notag \\
&&\times \left\{ \mathrm{Tr}\left[ \gamma _{5}\widetilde{S}_{s}^{b^{\prime
}b}(-x)\gamma _{5}S_{u}^{a^{\prime }a}(-x)\right] +\mathrm{Tr}\left[ \gamma
_{5}\widetilde{S}_{s}^{a^{\prime }b}(-x)\right. \right.  \notag \\
&&\left. \times \gamma _{5}S_{u}^{b^{\prime }a}(-x)\right] +\mathrm{Tr}\left[
\gamma _{5}\widetilde{S}_{s}^{b^{\prime }a}(-x)\gamma _{5}S_{u}^{a^{\prime
}b}(-x)\right]  \notag \\
&&\left. +\mathrm{Tr}\left[ \gamma _{5}\widetilde{S}_{s}^{a^{\prime
}a}(-x)\gamma _{5}S_{u}^{b^{\prime }b}(-x)\right] \right\}.  \label{eq;OPE2}
\end{eqnarray}%
The mass $\widetilde{m}$ and coupling $\widetilde{f}$ of the tetraquark $%
Z_{b:\overline{s}}^{0}$ can be found from Eqs.\ (\ref{eq:Mass}) and (\ref%
{eq:Coupl}) after replacing $\rho ^{\mathrm{OPE}}(s)$ by a relevant spectral
density $\widetilde{\rho }^{\mathrm{OPE}}(s)$ and using $\widetilde{\mathcal{%
M}}=m_{b}+m_{c}+m_{s}$ instead of $\mathcal{M}$. Predictions for $\widetilde{%
m}$ and $\widetilde{f}$ read
\begin{eqnarray}
\widetilde{m} &=&(6830\pm 160)~\mathrm{MeV},  \notag \\
\widetilde{f} &=&(7.1\pm 1.8)\times 10^{-3}~\mathrm{GeV}^{4}.
\label{eq:CMass}
\end{eqnarray}%
The $\widetilde{m}$ and $\widetilde{f}$ are extracted using the following
regions for the parameters $M^{2}$ and $s_{0}$%
\begin{equation}
M^{2}\in \lbrack 5.5,6.5]~\mathrm{GeV}^{2},\ s_{0}\in \lbrack 53,55]~\mathrm{%
GeV}^{2}.  \label{eq:Result2}
\end{equation}%
These working windows meet standard requirements of the sum rule
computations which have been discussed above. In fact, since at $M^{2}=5.5\
\mathrm{GeV}^{2}$ the ratio $R$ is equal to $0.008$, the convergence of the
obtained sum rules is guaranteed. The pole contribution at maximum of the
Borel parameter $M^{2}=6.5\ \mathrm{GeV}^{2}$ amounts to $\mathrm{PC}=0.25$,
which is in accord with the restriction (\ref{eq:PC}), and reaches $\mathrm{%
PC}=0.64$ at $M^{2}=5.5\ \mathrm{GeV}^{2}$. Theoretical uncertainties of
calculations for the mass $\pm 2.3\%$ are considerably smaller than
ambiguities of the coupling $\pm 25\%$ due to reasons explained above. In
Fig.\ \ref{fig:MassZ} we depict our prediction for the mass of the
tetraquark $Z_{b:\overline{s}}^{0}$ and show its dependence on $M^{2}$ and $%
s_{0}$.

In the framework of the QCD sum rule method the mass of scalar tetraquark $bb%
\overline{q}\overline{s}$ was evaluated in Ref. \cite{Du:2012wp}.
Computations there were carried out using different interpolating currents
and by taking into account nonperturbative terms up to dimension $8$.
Predictions obtained in Ref. \cite{Du:2012wp} $m=(10.2~\pm 0.3)~\mathrm{GeV}$
and $m=(10.3~\pm 0.3)~\mathrm{GeV}$ confirm a stable nature of this exotic
meson, and are very close to our result.


\section{Semileptonic decay $T_{b:\overline{s}}^{-}\rightarrow Z_{b:%
\overline{s}}^{0}l\overline{\protect\nu }_{l}$}

\label{sec:Decays1}
Our result $m=(10250~\pm 270)~\mathrm{MeV}$ for the mass of the tetraquark $%
T_{b:\overline{s}}^{-}$ demonstrates its stability against the strong and
electromagnetic decays to final states $B^{-}\overline{B}_{s}^{0}$ and $B^{-}%
\overline{B}_{s1}(5830)\gamma $, respectively. In fact, the central value of
the mass $m=10250~\mathrm{MeV}$ is $396~\mathrm{MeV}$ lower than the
threshold for strong decay to the conventional mesons $B^{-}\overline{B}%
_{s}^{0}$. Even its maximal value $10520~\mathrm{MeV}$ obtained by taking
into account uncertainties of the method is $126~\mathrm{MeV}$ below this
limit. Because the threshold $11108~\mathrm{MeV}$ for electromagnetic
dissociation of the $T_{b:\overline{s}}^{-}$ is considerably higher than $m$
the similar arguments hold for the corresponding process as well.

Therefore, the full width and lifetime of $T_{b:\overline{s}}^{-}$ are
determined by its weak transitions. In this section we concentrate on the
dominant semileptonic decay $T_{b:\overline{s}}^{-}\rightarrow Z_{b:%
\overline{s}}^{0}l\overline{\nu }_{l}$, which is depicted in Fig.\ \ref%
{fig:Decay1}. It is clear, that due to large mass difference\ $m-\widetilde{m%
}\approx 3420~\mathrm{MeV}$ decays $T_{b:\overline{s}}^{-}\rightarrow Z_{b:%
\overline{s}}^{0}l\overline{\nu }_{l}$ are kinematically allowed for all
lepton species $l=e,\ \mu $ and $\tau $. We do not consider processes
triggered by $b\rightarrow W^{-}u$, because they are suppressed relative to
dominant ones by a factor $|V_{bu}|^{2}/|V_{bc}|^{2}$ $\simeq 0.01$.

The transition $b\rightarrow W^{-}c$ at the tree-level can be described by
means of the effective Hamiltonian
\begin{equation}
\mathcal{H}^{\mathrm{eff}}=\frac{G_{F}}{\sqrt{2}}V_{bc}\overline{c}\gamma
_{\mu }(1-\gamma _{5})b\overline{l}\gamma ^{\mu }(1-\gamma _{5})\nu _{l},
\label{eq:EffecH}
\end{equation}%
where $G_{F}$ and $V_{bc}$ are the Fermi coupling constant and the relevant
CKM matrix element, respectively:%
\begin{eqnarray}
G_{F} &=&1.16637\times 10^{-5}~\mathrm{GeV}^{-2},  \notag \\
|V_{bc}| &=&(42.2\pm 0.08)\times 10^{-3}.
\end{eqnarray}%
After sandwiching $\mathcal{H}^{\mathrm{eff}}$ between the initial and final
tetraquark fields, and removing a leptonic part from an obtained expression,
we get the matrix element of the current%
\begin{equation}
J_{\mu }^{\mathrm{tr}}=\overline{c}\gamma _{\mu }(1-\gamma _{5})b.
\label{eq:TrCurr}
\end{equation}%
The latter can be written down using the form factors $G_{i}(q^{2})$ ($i=1,2$%
) which parametrize the long-distance dynamics of the weak transition. In
terms of $G_{1(2)}(q^{2})$ the matrix element of the current $J_{\mu }^{%
\mathrm{tr}}$ has the form
\begin{eqnarray}
\langle Z_{b:\overline{s}}^{0}(p^{\prime })|J_{\mu }^{\mathrm{tr}}|T_{b:%
\overline{s}}^{-}(p)\rangle &=&G_{1}(q^{2})P_{\mu }+G_{2}(q^{2})q_{\mu },
\notag \\
&&  \label{eq:Vertex1}
\end{eqnarray}%
where $p$ and $p^{\prime }$ are the momenta of the initial and final
tetraquarks, respectively. Here we also introduce variables $P_{\mu }=p_{\mu
}^{\prime }+p_{\mu }$ and $q_{\mu }=p_{\mu }-p_{\mu }^{\prime }$. The $%
q_{\mu }$ is the momentum transferred to the leptons, and $q^{2}$ changes
within the limits $m_{l}^{2}\leq q^{2}\leq (m-\widetilde{m})^{2}$ with $%
m_{l} $ being the mass of a lepton $l$.
\begin{figure}[h!]
\begin{center}
\includegraphics[totalheight=6cm,width=8cm]{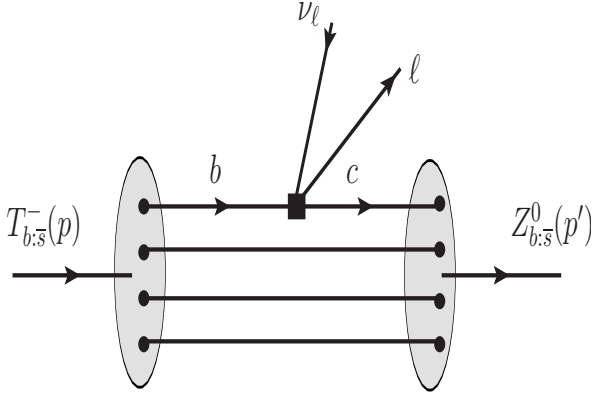}
\end{center}
\caption{The Feynman diagram for the semileptonic decay $T_{b:\overline{s}%
}^{-}\rightarrow Z_{b:\overline{s}}^{0}l\overline{\protect\nu }_{l}$. The
black square denotes the effective weak vertex.}
\label{fig:Decay1}
\end{figure}

To derive the sum rules for the form factors $G_{1(2)}(q^{2})$, we begin
from analysis of the three-point correlation function
\begin{eqnarray}
\Pi _{\mu }(p,p^{\prime }) &=&i^{2}\int d^{4}xd^{4}ye^{i(p^{\prime }y-px)}
\notag \\
&&\times \langle 0|\mathcal{T}\{\widetilde{J}(y)J_{\mu }^{\mathrm{tr}%
}(0)J^{\dagger }(x)\}|0\rangle .  \label{eq:CF2}
\end{eqnarray}

In accordance with standard prescriptions, we write the correlation function
$\Pi _{\mu }(p,p^{\prime })$ using the spectroscopic parameters of the
tetraquarks, and get the physical side of the sum rule $\Pi _{\mu }^{\mathrm{%
Phys}}(p,p^{\prime })$. The function $\Pi _{\mu }^{\mathrm{Phys}%
}(p,p^{\prime })$ can be presented in the following form%
\begin{eqnarray}
&&\Pi _{\mu }^{\mathrm{Phys}}(p,p^{\prime })=\frac{\langle 0|\widetilde{J}%
|Z_{b:\overline{s}}^{0}(p^{\prime })\rangle \langle Z_{b:\overline{s}%
}^{0}(p^{\prime })|J_{\mu }^{\mathrm{tr}}|T_{b:\overline{s}}^{-}(p)\rangle }{%
(p^{2}-m^{2})(p^{\prime 2}-\widetilde{m}^{2})}  \notag \\
&&\times \langle T_{b:\overline{s}}^{-}(p)|J^{\dagger }|0\rangle +\cdots,
\label{eq:CF3}
\end{eqnarray}%
where the contribution of the ground-state particles is shown explicitly,
whereas effects of excited resonances and continuum states are denoted by
dots.

The phenomenological side of the sum rules can be detailed by expressing the
matrix elements in terms of the tetraquarks' mass and coupling, and weak
transition form factors. For these purposes, we use Eqs.\ (\ref{eq:ME1}) and
(\ref{eq:Vertex1}), and employ the matrix element of the state $Z_{b:%
\overline{s}}^{0}$
\begin{equation}
\langle 0|\widetilde{J}|Z_{b:\overline{s}}^{0}(p^{\prime })\rangle =%
\widetilde{f}\widetilde{m}.  \label{eq:ME3}
\end{equation}%
Then it is not difficult to find that
\begin{eqnarray}
\Pi _{\mu }^{\mathrm{Phys}}(p,p^{\prime }) &=&\frac{fm\widetilde{f}%
\widetilde{m}}{(p^{2}-m^{2})(p^{\prime 2}-\widetilde{m}^{2})}  \notag \\
&&\times \left[ G_{1}(q^{2})P_{\mu }+G_{2}(q^{2})q_{\mu }\right] +\cdots.
\label{eq:Phys1}
\end{eqnarray}

We determine $\Pi _{\mu }(p,p^{\prime })$ also by utilizing the
interpolating currents and quark propagators, which lead to QCD side of the
sum rules
\begin{eqnarray}
&&\Pi _{\mu }^{\mathrm{OPE}}(p,p^{\prime })=i^{2}\int
d^{4}xd^{4}ye^{i(p^{\prime }y-px)}\left( \mathrm{Tr}\left[ \gamma _{5}%
\widetilde{S}_{s}^{b^{\prime }b}(x-y)\right. \right.  \notag \\
&&\left. \left. \times \gamma _{5}S_{u}^{a^{\prime }a}(x-y)\right] +\mathrm{%
Tr}\left[ \gamma _{5}\widetilde{S}_{s}^{b^{\prime }a}(x-y)\gamma _{\nu
}S_{u}^{a^{\prime }b}(x-y)\right] \right)  \notag \\
&&\times \left( \mathrm{Tr}\left[ \gamma _{5}\widetilde{S}_{b}^{aa^{\prime
}}(y-x)\gamma _{5}S_{c}^{bi}(y)\gamma _{\mu }(1-\gamma
_{5})S_{b}^{ib^{\prime }}(-x)\right] \right.  \notag \\
&&\left. +\mathrm{Tr}\left[ \gamma _{5}\widetilde{S}_{b}^{ia^{\prime
}}(-x)(1-\gamma _{5})\gamma _{\mu }\widetilde{S}_{c}^{bi}(y)\gamma
_{5}S_{b}^{ab^{\prime }}(y-x)\right] \right).  \notag \\
&&  \label{eq:QCD2}
\end{eqnarray}

One can obtain the sum rules for the form factors $G_{1(2)}(q^{2})$ by
equating invariant amplitudes corresponding to structures $P_{\mu }$ and $%
q_{\mu }$ from $\Pi _{\mu }^{\mathrm{Phys}}(p,p^{\prime })$ and $\Pi _{\mu
}^{\mathrm{OPE}}(p,p^{\prime })$. It is known that, these invariant
amplitudes depend on $p^{2}$ and $p^{\prime 2}$, and therefore in order to
suppress contributions of higher resonances and continuum states we have to
apply the double Borel transformation over these variables. As a result, the
final expressions contain a set of Borel parameters $\mathbf{M}%
^{2}=(M_{1}^{2},\ M_{2}^{2})$. The continuum subtraction should be carried
out in two channels which generates a dependence on the threshold parameters
$\mathbf{s}_{0}=(s_{0},\ s_{0}^{\prime })$.

These operations lead to the sum rules
\begin{eqnarray}
&&G_{i}(\mathbf{M}^{2},\mathbf{s}_{0},q^{2})=\frac{1}{fm\widetilde{f}%
\widetilde{m}}\int_{\mathcal{M}^{2}}^{s_{0}}ds  \notag \\
&&\times \int_{\widetilde{\mathcal{M}}^{2}}^{s_{0}^{\prime }}ds^{\prime
}\rho _{i}(s,s^{\prime },q^{2})e^{(m^{2}-s)/M_{1}^{2}}e^{(\widetilde{m}%
^{2}-s^{\prime })/M_{2}^{2}},  \notag \\
&&  \label{eq:SR}
\end{eqnarray}%
where $\rho _{1(2)}(s,s^{\prime },q^{2})$ are the spectral densities
calculated as the imaginary part of the correlation function $\Pi _{\mu }^{%
\mathrm{OPE}}(p,p^{\prime })$ with dimension-7 accuracy. The first pair of
parameters $(M_{1}^{2},s_{0})$ in Eq.\ (\ref{eq:SR}) is related to the
initial state $T_{b:\overline{s}}^{-}$, whereas the second set $%
(M_{2}^{2},s_{0}^{\prime })$ corresponds to the final particle $Z_{b:%
\overline{s}}^{0}$. Explicit expressions of $\rho _{1(2)}(s,s^{\prime
},q^{2}) $ are rather cumbersome, therefore we do not provide them here.

\begin{figure}[h]
\includegraphics[width=8.8cm]{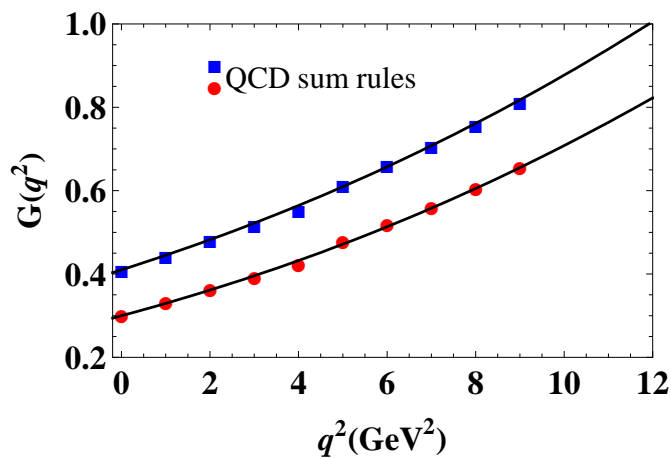}
\caption{Predictions for the form factors $|G_{1}(q^{2})|$ (the lower red
circles) and $G_{2}(q^{2})$ (the upper blue squares). The lines are fit
functions $|F_{1}(q^{2})|$ and $F_{2}(q^{2})$, respectively.}
\label{fig:FFG}
\end{figure}

In numerical computations of $G_{1(2)}(q^{2})$ the working regions for the
parameters $\mathbf{M}^{2}$ and $\mathbf{s}_{0}$ are chosen exactly as in
the corresponding mass calculations. Values of the vacuum condensates are
collected in Eq.\ (\ref{eq:Parameters}), whereas the masses and couplings of
the tetraquarks $T_{b:\overline{s}}^{-}$ and $Z_{b:\overline{s}}^{0}$ have
been calculated in the present work and written down in Eqs.\ (\ref%
{eq:Result1}) and (\ref{eq:Result2}), respectively. Obtained sum rule
predictions for the form factors $G_{1}(q^{2})$ and $G_{2}(q^{2})$ are shown
in Fig.\ \ref{fig:FFG}.

The sum rules give reliable results for $G_{1(2)}(q^{2})$ in the region $%
m_{l}^{2}\leq q^{2}\leq 9~\mathrm{GeV}^{2}$. But this is not enough to
calculate the partial width of the decay $T_{b:\overline{s}}^{-}\rightarrow
Z_{b:\overline{s}}^{0}l\overline{\nu }_{l}$ under analysis. Indeed, the form
factors determine the differential decay rate $d\Gamma /dq^{2}$ of the
process through the following expression%
\begin{eqnarray}
&&\frac{d\Gamma }{dq^{2}}=\frac{G_{F}^{2}|V_{bc}|^{2}}{64\pi ^{3}m^{3}}%
\lambda \left( m^{2},\widetilde{m}^{2},q^{2}\right) \left( \frac{%
q^{2}-m_{l}^{2}}{q^{2}}\right) ^{2}  \notag \\
\times &&\left\{ (2q^{2}+m_{l}^{2})\left[ G_{1}^{2}(q^{2})\left( \frac{q^{2}%
}{2}-m^{2}-\widetilde{m}^{2}\right) \right. \right.  \notag \\
&&\left. -G_{2}^{2}(q^{2})\frac{q^{2}}{2}+(\widetilde{m}%
^{2}-m^{2})G_{1}(q^{2})G_{2}(q^{2})\right]  \notag \\
&&\left. +\frac{q^{2}+m_{l}^{2}}{q^{2}}\left[ G_{1}(q^{2})(m^{2}-\widetilde{m%
}^{2})+G_{2}(q^{2})q^{2}\right] ^{2}\right\},  \notag \\
&&  \label{eq:Drate}
\end{eqnarray}%
where%
\begin{eqnarray}
&&\lambda \left( m^{2},\widetilde{m}^{2},q^{2}\right) =\left[ m^{4}+%
\widetilde{m}^{4}+q^{4}\right.  \notag \\
&&\left. -2\left( m^{2}\widetilde{m}^{2}+m^{2}q^{2}+\widetilde{m}%
^{2}q^{2}\right) \right] ^{1/2}.
\end{eqnarray}%
To find the partial width of the semileptonic decay, $d\Gamma /dq^{2}$
should be integrated over $q^{2}$ in the limits $m_{l}^{2}\leq q^{2}\leq (m-%
\widetilde{m})^{2}$. But the region $m_{l}^{2}\leq q^{2}\leq 11.7~\mathrm{GeV%
}^{2}$ is wider than a domain where the sum rules lead to strong
predictions. This problem can be solved by introducing model functions $%
F_{i}(q^{2})$ which at the momentum transfers $q^{2}$ accessible for the sum
rule computations coincide with $G_{i}(q^{2})$, but can be extrapolated to
the whole integration region. These functions should have a simple form and
be suitable to perform integrations over $q^{2}$.

To this end, we use the functions of the form
\begin{equation}
F_{i}(q^{2})=F_{0}^{i}\exp \left[ c_{1}^{i}\frac{q^{2}}{m^{2}}%
+c_{2}^{i}\left( \frac{q^{2}}{m^{2}}\right) ^{2}\right],
\label{eq:FFunctions}
\end{equation}%
where $F_{0}^{i},~c_{1}^{i}$, and $c_{2}^{i}$ are constants which have to be
fixed by comparing $F_{i}(q^{2})$ and $G_{i}(q^{2})$ at common regions of
validity. Numerical analysis allows us to fix
\begin{eqnarray}
F_{0}^{1} &=&-0.30,~c_{1}^{1}=9.98,\ c_{2}^{1}=-10.07,  \notag \\
F_{0}^{2} &=&0.41,~c_{1}^{2}=8.67,\ c_{2}^{2}=-7.15.
\label{eq:Fitparameters}
\end{eqnarray}%
The functions $F_{i}(q^{2})$ are plotted in Fig. \ref{fig:FFG}, where one
can see their nice agreement with the sum rule predictions.

Other input information to calculate the partial width of the process $T_{b:%
\overline{s}}^{-}\rightarrow Z_{b:\overline{s}}^{0}l\overline{\nu }_{l}$,
namely the masses of the leptons $m_{e}=0.511~\mathrm{MeV}$, $m_{\mu
}=105.658~\mathrm{MeV}$, and $m_{\tau }=(1776.82~\pm 0.16)~\mathrm{MeV}$ are
borrowed from Ref.\ \cite{Tanabashi:2018oca}.

Our results for the partial widths of the semileptonic decay channels are
presented below:
\begin{eqnarray}
\Gamma (T_{b:\overline{s}}^{-} &\rightarrow &Z_{b:\overline{s}}^{0}e^{-}%
\overline{\nu }_{e})=(6.16\pm 1.74)\times 10^{-10}~\mathrm{MeV},  \notag \\
\Gamma (T_{b:\overline{s}}^{-} &\rightarrow &Z_{b:\overline{s}}^{0}\mu ^{-}%
\overline{\nu }_{\mu })=(6.15\pm 1.74)\times 10^{-10}~\mathrm{MeV},  \notag
\\
\Gamma (T_{b:\overline{s}}^{-} &\rightarrow &Z_{b:\overline{s}}^{0}\tau ^{-}%
\overline{\nu }_{\tau })=(2.85\pm 0.81)\times 10^{-10}~\mathrm{MeV}.  \notag
\\
&&  \label{eq:Results}
\end{eqnarray}%
As we shall see below, the semileptonic decays $T_{b:\overline{s}%
}^{-}\rightarrow Z_{b:\overline{s}}^{0}l\overline{\nu }_{l}$ establish an
essential part of the full width of $T_{b:\overline{s}}^{-}$.


\section{ Nonleptonic decays $T_{b:\overline{s}}^{-}\rightarrow Z_{b:%
\overline{s}}^{0}\protect\pi ^{-}(K^{-},\ D^{-},\ D_{s}^{-})$}

\label{sec:Decays2}

In this section, we investigate the nonleptonic weak decays $T_{b:\overline{s%
}}^{-}\rightarrow Z_{b:\overline{s}}^{0}\pi ^{-}(K^{-},\ D^{-},\ D_{s}^{-})$
of the tetraquark $T_{b:\overline{s}}^{-}$ in the framework of the QCD
factorization method, which allows us to calculate partial widths of these
processes. This approach was applied to investigate nonleptonic decays of
the conventional mesons \cite{Beneke:1999br,Beneke:2000ry}, and used to
study nonleptonic decays of the scalar and axial-vector tetraquarks $Z_{bc;%
\overline{u}\overline{d}}^{0}$ and $T_{bc;\overline{u}\overline{d}}^{0}$ in
Refs.\ \cite{Sundu:2019feu,Agaev:2019kkz}, respectively.

Here, we consider in a detailed form the decay $T_{b:\overline{s}%
}^{-}\rightarrow Z_{b:\overline{s}}^{0}\pi ^{-}$ shown in Fig.\ \ref%
{fig:Decay2}, and write down final predictions for remaining channels. At
the quark level, the effective Hamiltonian for this decay is given by the
expression
\begin{equation}
\mathcal{H}_{\mathrm{n.-lep}}^{\mathrm{eff}}=\frac{G_{F}}{\sqrt{2}}%
V_{bc}V_{ud}^{\ast }\left[ c_{1}(\mu )Q_{1}+c_{2}(\mu )Q_{2}\right],
\label{eq:EffHam}
\end{equation}%
where%
\begin{eqnarray}
Q_{1} &=&\left( \overline{d}_{i}u_{i}\right) _{\mathrm{V-A}}\left( \overline{%
c}_{j}b_{j}\right) _{\mathrm{V-A}},  \notag \\
Q_{2} &=&\left( \overline{d}_{i}u_{j}\right) _{\mathrm{V-A}}\left( \overline{%
c}_{j}b_{i}\right) _{\mathrm{V-A}},  \label{eq:Operators}
\end{eqnarray}%
and $i$ , $j$ are the color indices, and notation $\left( \overline{q}%
_{1}q_{2}\right) _{\mathrm{V-A}}$ means
\begin{equation}
\left( \overline{q}_{1}q_{2}\right) _{\mathrm{V-A}}=\overline{q}_{1}\gamma
_{\mu }(1-\gamma _{5})q_{2}.  \label{eq:Not}
\end{equation}%
The short-distance Wilson coefficients $c_{1}(\mu )$ and $c_{2}(\mu )$ are
given at the factorization scale $\mu $.

\begin{figure}[h!]
\begin{center}
\includegraphics[totalheight=6cm,width=8cm]{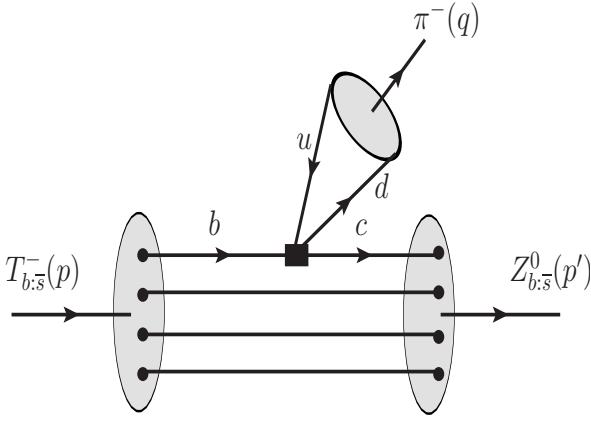}
\end{center}
\caption{The same as in Fig.\ 3, but for the nonleptonic decay $T_{b:%
\overline{s}}^{-}\rightarrow Z_{b:\overline{s}}^{0}\protect\pi ^{-}$}
\label{fig:Decay2}
\end{figure}

The amplitude of the decay $T_{b:\overline{s}}^{-}\rightarrow Z_{b:\overline{%
s}}^{0}\pi ^{-}$ can be presented in the factorized form%
\begin{eqnarray}
\mathcal{A} &=&\frac{G_{F}}{\sqrt{2}}V_{bc}V_{ud}^{\ast }a_{1}(\mu )\langle
\pi ^{-}(q)|\left( \overline{d}_{i}u_{i}\right) _{\mathrm{V-A}}|0\rangle
\notag \\
&&\times \langle Z_{b:\overline{s}}^{0}(p^{\prime })|\left( \overline{c}%
_{j}b_{j}\right) _{\mathrm{V-A}}|T_{b:\overline{s}}^{-}(p)\rangle,
\label{eq:Amplitude}
\end{eqnarray}%
where
\begin{equation}
a_{1}(\mu )=c_{1}(\mu )+\frac{1}{N_{c}}c_{2}(\mu ),
\end{equation}%
and $N_{c}=3$ is the number of quark colors. The amplitude $\mathcal{A}$
describes the process in which the pion $\pi ^{-}$ is generated directly
from the color-singlet current $\left( \overline{d}_{i}u_{i}\right) _{%
\mathrm{V-A}}$. The matrix element $\langle Z_{b:\overline{s}}^{0}(p^{\prime
})|\left( \overline{c}_{j}b_{j}\right) _{\mathrm{V-A}}|T_{b:\overline{s}%
}^{-}(p)\rangle $ has been introduced in Eq.\ (\ref{eq:Vertex1}), while the
matrix element of the pion is given by the expression%
\begin{equation}
\langle \pi ^{-}(q)|\left( \overline{d}_{i}u_{i}\right) _{\mathrm{V-A}%
}|0\rangle =if_{\pi }q_{\mu },  \label{eq:ME4}
\end{equation}%
where $f_{\pi }$ is the decay constant of $\pi $.

Then, it is not difficult to see that $\mathcal{A}$ takes the form%
\begin{eqnarray}
\mathcal{A} &=&i\frac{G_{F}}{\sqrt{2}}f_{\pi }V_{bc}V_{ud}^{\ast }a_{1}(\mu )%
\left[ G_{1}(q^{2})Pq+G_{2}(q^{2})q^{2}\right].  \notag \\
&&  \label{eq:Amplitude2}
\end{eqnarray}%
The partial width of this process is determined by the simple expression%
\begin{eqnarray}
&&\Gamma (T_{b:\overline{s}}^{-}\rightarrow Z_{b:\overline{s}}^{0}\pi ^{-})=%
\frac{G_{F}^{2}f_{\pi }^{2}|V_{bc}|^{2}|V_{ud}|^{2}}{32\pi m^{3}}%
a_{1}^{2}(\mu )  \notag \\
&&\times \lambda \left( m^{2},\widetilde{m}^{2},m_{\pi }^{2}\right) \left[
G_{1}(m^{2}-\widetilde{m}^{2})+G_{2}m_{\pi }^{2}\right] ^{2},
\label{eq:NLDW}
\end{eqnarray}%
where the weak form factors $G_{1(2)}(q^{2})$ are computed at $q^{2}=m_{\pi
}^{2}$. The similar analysis can be performed for the decay modes $T_{b:%
\overline{s}}^{-}\rightarrow Z_{b:\overline{s}}^{0}K^{-}(D^{-},\ D_{s}^{-})$
as well. The partial width of these channels can be obtained from Eq.\ (\ref%
{eq:NLDW}) by replacing ($m_{\pi },f_{\pi }$) with the spectroscopic
parameters of the mesons $K$, $D$, and $D_{s}$, and implementing
substitutions $|V_{ud}|\rightarrow |V_{us}|$, $|V_{cd}|$, and $|V_{cs}|$,
respectively.

The masses and decay constants of the final-state pseudoscalar mesons, as
well as values of the CKM matrix elements used in computations are collected
in Table\ \ref{tab:MesonPar}. The Wilson coefficients $c_{1}(m_{b}),\ $and $%
c_{2}(m_{b})$ with next-to-leading order QCD corrections can be found in
Refs.\ \cite{Buras:1992zv,Ciuchini:1993vr,Buchalla:1995vs}
\begin{equation}
c_{1}(m_{b})=1.117,\ c_{2}(m_{b})=-0.257.  \label{eq:WCoeff}
\end{equation}

\begin{table}[tbp]
\begin{tabular}{|c|c|}
\hline\hline
Quantity & Value \\ \hline\hline
$m_{\pi} $ & $139.570~\mathrm{MeV}$ \\
$m_{K}$ & $(493.677\pm 0.016)~\mathrm{MeV}$ \\
$m_{D}$ & $(1869.61 \pm 0.10)~\mathrm{MeV}$ \\
$m_{D_s}$ & $(1968.30\pm 0.11)~\mathrm{MeV}$ \\
$f_{\pi }$ & $131~\mathrm{MeV}$ \\
$f_{K}$ & $(155.72\pm 0.51)~\mathrm{MeV}$ \\
$f_{D}$ & $(203.7 \pm 4.7)~\mathrm{MeV}$ \\
$f_{D_s}$ & $(257.8 \pm 4.1)~\mathrm{MeV}$ \\
$|V_{ud}|$ & $0.97420\pm 0.00021$ \\
$|V_{us}|$ & $0.2243\pm 0.0005$ \\
$|V_{cd}|$ & $0.218\pm 0.004$ \\
$|V_{cs}|$ & $0.997\pm 0.017$ \\ \hline\hline
\end{tabular}%
\caption{Spectroscopic parameters of the final-state pseudoscalar mesons,
and the relevant CKM matrix elements. }
\label{tab:MesonPar}
\end{table}

For the decay $T_{b:\overline{s}}^{-}\rightarrow Z_{b:\overline{s}}^{0}\pi
^{-}$ calculations lead to the result
\begin{eqnarray}
\Gamma (T_{b:\overline{s}}^{-} &\rightarrow &Z_{b:\overline{s}}^{0}\pi
^{-})=\left( 6.97\pm 1.99\right) \times 10^{-13}~\mathrm{MeV}.  \notag \\
&&  \label{eq:NLDW1}
\end{eqnarray}
For the remaining nonleptonic decays of the tetraquark $T_{b:\overline{s}%
}^{-}$, we get
\begin{eqnarray}
\Gamma (T_{b:\overline{s}}^{-} &\rightarrow &Z_{b:\overline{s}%
}^{0}K^{-})=\left( 5.33\pm 1.47\right) \times 10^{-14}~\mathrm{MeV},  \notag
\\
\Gamma (T_{b:\overline{s}}^{-} &\rightarrow &Z_{b:\overline{s}%
}^{0}D^{-})=\left( 1.13\pm 0.31\right) \times 10^{-13}~\mathrm{MeV},  \notag
\\
\Gamma (T_{b:\overline{s}}^{-} &\rightarrow &Z_{b:\overline{s}%
}^{0}D_{s}^{-})=\left( 3.88\pm 1.01\right) \times 10^{-12}~\mathrm{MeV}.
\notag \\
&&  \label{eq:NLDW2}
\end{eqnarray}%
It is seen that partial widths of the nonleptonic decays are negligibly
smaller than widths of the semileptonic decays. Only widths of the processes
$T_{b:\overline{s}}^{-}\rightarrow Z_{b:\overline{s}}^{0}\pi ^{-}$ and $T_{b:%
\overline{s}}^{-}\rightarrow Z_{b:\overline{s}}^{0}D_{s}^{-}$ affect the
final result for $\Gamma _{\mathrm{full}}$.

Collected information on the partial widths of the weak decays of the
tetraquark $T_{b:\overline{s}}^{-}$ allow us to find its full width and mean
lifetime:

\begin{eqnarray}
\Gamma _{\mathrm{full}} &=&(15.21\pm 2.59)\times 10^{-10}~\mathrm{MeV},
\notag \\
\tau &=&4.33_{-0.63}^{+0.89}\times 10^{-13}~\mathrm{s}.  \label{eq:WL}
\end{eqnarray}%
Predictions for $\Gamma _{\mathrm{full}}$ and $\tau $ are the main results of
the present work.


\section{Discussion and concluding notes}

\label{sec:Disc}

In this article we have evaluated the mass and coupling of the scalar
tetraquark $T_{b:\overline{s}}^{-}$. Our analysis has proved that the exotic
meson $T_{b:\overline{s}}^{-}$ composed of the heavy diquark $bb$ and light
antidiquark $\overline{u}\overline{s}$ is the strong- and
electromagnetic-interaction stable state, and dissociates to conventional
mesons only through the weak decays. This fact places it to a list of stable
axial-vector $T_{bb;\overline{u}\overline{d}}^{-}$ , and scalar $Z_{bc;%
\overline{u}\overline{d}}^{0}$ and $T_{bs;\overline{u}\overline{d}}^{-}$
tetraquarks.
\begin{table}[tbp]
\begin{tabular}{|c|c|}
\hline\hline
Channels & $\mathcal{BR}$ \\ \hline\hline
$\Gamma (T_{b:\overline{s}}^{-} \to Z_{b:\overline{s}}^{0}e^{-}\overline{\nu
}_{e})$ & $0.404$ \\
$\Gamma (T_{b:\overline{s}}^{-} \to Z_{b:\overline{s}}^{0}\mu ^{-}\overline{%
\nu }_{\mu })$ & $0.404$ \\
$\Gamma (T_{b:\overline{s}}^{-} \to Z_{b:\overline{s}}^{0}\tau ^{-}\overline{%
\nu }_{\tau })$ & $0.187$ \\
$\Gamma (T_{b:\overline{s}}^{-} \to Z_{b:\overline{s}}^{0}\pi ^{-})$ & $%
4.58\times 10^{-4}$ \\
$\Gamma (T_{b:\overline{s}}^{-} \to Z_{b:\overline{s}}^{0}D_{s}^{-})$ & $%
2.55\times 10^{-3}$ \\ \hline\hline
\end{tabular}%
\caption{Dominant weak decay channels of the tetraquark $T_{b:\overline{s}%
}^{-}$, and corresponding branching ratios.}
\label{tab:BR}
\end{table}

We have investigated also the dominant weak decay modes of the $T_{b:%
\overline{s}}^{-}$, and computed their partial widths. These results have
allowed us to estimate the full width and mean lifetime of the $T_{b:%
\overline{s}}^{-}$. The collected information is enough to find the
branching ratios of the various decay modes as well (see Table\ \ref{tab:BR}%
). It is worth noting that only the semileptonic decays of the $T_{b:%
\overline{s}}^{-}$ play a dominant role in forming of $\Gamma _{\mathrm{full}%
}$.

The tetraquark $T_{b:\overline{s}}^{-}$ can be considered as "$s$" member of
a multiplet of the scalar $bb\overline{q}\overline{q}^{\prime }$ states with
$q$ being one of the light quarks. In Ref.\ \cite{Agaev:2018khe}, we studied
the stable axial-vector\ particle $T_{bb;\overline{u}\overline{d}}^{-}$. It
will be very interesting to investigate the scalar partner of $T_{bb;%
\overline{u}\overline{d}}^{-}$, as well as the axial-vector state $bc%
\overline{u}\overline{s}$ which may shed light on others members of scalar
and axial-vector multiplets $bb\overline{q}\overline{q}^{\prime }$.

We have computed the mass and coupling of the scalar tetraquark $Z_{b:%
\overline{s}}^{0}$: these parameters are required to explore the weak decays
of the $T_{b:\overline{s}}^{-}$. The state $Z_{b:\overline{s}}^{0}=bc%
\overline{u}\overline{s}$ belongs to a famous class of exotic mesons
composed of four different quarks \cite{Agaev:2016mjb}. A simple analysis
confirms that it is a strong-interaction stable particle. Indeed, the scalar
tetraquark $Z_{b:\overline{s}}^{0}$ in $S$-wave may decay to a pair of
pseudoscalar mesons $B^{-}D_{s}^{+}$ and $\overline{B}_{s}^{0}D^{0}$.
Thresholds for production of these pairs are $7248~\mathrm{MeV}$ and $7237~%
\mathrm{MeV}$, respectively. Because the maximum allowed value of the $Z_{b:%
\overline{s}}^{0}$ tetraquarks's mass is $\widetilde{m}=6990~\mathrm{MeV}$,
it is stable against these strong decays. The $\overline{u}\overline{d}$
member of the scalar multiplet $bc\overline{q}\overline{q}^{\prime }$ was
investigated in Ref.\ \cite{Sundu:2019feu}, in which it was found that this
particle is a strong- and electromagnetic-interaction stable state. It seems
scalar particles with such diquark-antidiquark structures are among real
candidates to stable four-quark compounds.

The revealed features of the $Z_{b:\overline{s}}^{0}$ determine a decay
pattern of the master particle $T_{b:\overline{s}}^{-}$. Indeed, the
tetraquark $Z_{b:\overline{s}}^{0}$ created at the first stage of the
decays, at the next step due to subprocesses $b\rightarrow W^{-}c$ and $%
c\rightarrow W^{+}s$ should have undergone weak transformations. Such cascade
picture of decays was encountered in theoretical investigations of other
tetraquarks \cite{Agaev:2018khe,Sundu:2019feu}, and studied in a detailed
form in Ref.\ \cite{Agaev:2019wkk}. Of course, there are nonleptonic decays
of $T_{b:\overline{s}}^{-}$ when it transforms to a pair of ordinary mesons
at the first phase of a weak process. A comprehensive analysis of the $T_{b:%
\overline{s}}^{-}$ tetraquark's decays will be finished in our forthcoming
publications.

\section*{ACKNOWLEDGMENTS}

The work of K.~A, B.~B., and H.~S was supported in part by the TUBITAK grant
under No: 119F050.

\appendix*

\begin{widetext}

\section{ The propagators $S_{q(Q)}(x)$ and invariant amplitude $\Pi
(M^{2},s_{0})$}

\renewcommand{\theequation}{\Alph{section}.\arabic{equation}} \label{sec:App}

In the present work we use the light quark propagator $S_{q}^{ab}(x)$ which
is given by the following formula
\begin{eqnarray}
&&S_{q}^{ab}(x)=i\delta _{ab}\frac{\slashed x}{2\pi ^{2}x^{4}}-\delta _{ab}%
\frac{m_{q}}{4\pi ^{2}x^{2}}-\delta _{ab}\frac{\langle \overline{q}q\rangle
}{12}+i\delta _{ab}\frac{\slashed xm_{q}\langle \overline{q}q\rangle }{48}%
-\delta _{ab}\frac{x^{2}}{192}\langle \overline{q}g_{s}\sigma Gq\rangle
\notag \\
&&+i\delta _{ab}\frac{x^{2}\slashed xm_{q}}{1152}\langle \overline{q}%
g_{s}\sigma Gq\rangle -i\frac{g_{s}G_{ab}^{\alpha \beta }}{32\pi ^{2}x^{2}}%
\left[ \slashed x{\sigma _{\alpha \beta }+\sigma _{\alpha \beta }}\slashed x%
\right] -i\delta _{ab}\frac{x^{2}\slashed xg_{s}^{2}\langle \overline{q}%
q\rangle ^{2}}{7776}  \notag \\
&&-\delta _{ab}\frac{x^{4}\langle \overline{q}q\rangle \langle
g_{s}^{2}G^{2}\rangle }{27648}+\cdots.
\end{eqnarray}%
For the heavy quarks $Q$ we utilize the propagator $S_{Q}^{ab}(x)$
\begin{eqnarray}
&&S_{Q}^{ab}(x)=i\int \frac{d^{4}k}{(2\pi )^{4}}e^{-ikx}\Bigg \{\frac{\delta
_{ab}\left( {\slashed k}+m_{Q}\right) }{k^{2}-m_{Q}^{2}}-\frac{%
g_{s}G_{ab}^{\alpha \beta }}{4}\frac{\sigma _{\alpha \beta }\left( {\slashed %
k}+m_{Q}\right) +\left( {\slashed k}+m_{Q}\right) \sigma _{\alpha \beta }}{%
(k^{2}-m_{Q}^{2})^{2}}  \notag \\
&&+\frac{g_{s}^{2}G^{2}}{12}\delta _{ab}m_{Q}\frac{k^{2}+m_{Q}{\slashed k}}{%
(k^{2}-m_{Q}^{2})^{4}}+\frac{g_{s}^{3}G^{3}}{48}\delta _{ab}\frac{\left( {%
\slashed k}+m_{Q}\right) }{(k^{2}-m_{Q}^{2})^{6}}\left[ {\slashed k}\left(
k^{2}-3m_{Q}^{2}\right) +2m_{Q}\left( 2k^{2}-m_{Q}^{2}\right) \right] \left(
{\slashed k}+m_{Q}\right) +\cdots \Bigg \}.  \notag \\
&&
\end{eqnarray}

Above, we have introduced the notations
\begin{equation}
G_{ab}^{\alpha \beta }\equiv G_{A}^{\alpha \beta }t_{ab}^{A},\ \
G^{2}=G_{\alpha \beta }^{A}G_{A}^{\alpha \beta },\ G^{3}=f^{ABC}G_{\alpha
\beta }^{A}G^{B\beta \delta }G_{\delta }^{C\alpha },
\end{equation}%
where $G_{A}^{\alpha \beta }$ is the gluon field strength tensor, $%
t^{A}=\lambda ^{A}/2$ with $\lambda ^{A}$ being the Gell-Mann matrices, $%
f^{ABC}$ are the structure constants of the color group $SU_{c}(3),$ and $%
A,B,C=1,2,\ldots 8$.

The invariant amplitude $\Pi ^{\mathrm{OPE}}(p^{2})$ used for calculation of
the mass and coupling of the tetraquark $T_{b:\overline{s}}^{-}$ after the
Borel transformation and subtraction procedures takes the following form%
\begin{equation}
\Pi (M^{2},s_{0})=\int_{\mathcal{M}^{2}}^{s_{0}}ds\rho ^{\mathrm{OPE}%
}(s)e^{-s/M^{2}}+\Pi (M^{2}),
\end{equation}%
where
\begin{equation}
\rho ^{\mathrm{OPE}}(s)=\rho ^{\mathrm{pert.}}(s)+\sum_{N=3}^{8}\rho ^{%
\mathrm{DimN}}(s),\ \ \Pi (M^{2})=\sum_{N=6}^{10}\Pi ^{\mathrm{DimN}}(M^{2}).
\label{eq:A1}
\end{equation}%
Components of the spectral density are given by the formulas%
\begin{equation}
\rho (s)=\int_{0}^{1}d\alpha \int_{0}^{1-a}d\beta \rho (s,\alpha ,\beta ),\
\ \rho (s)=\int_{0}^{1}d\alpha \rho (s,\alpha ),  \label{eq:A2}
\end{equation}%
depending on whether $\rho (s,\alpha ,\beta )$ is a function of $\alpha $
and $\beta $ or only $\alpha$. The same is true also for terms $\Pi (M^{2})$%
, i.e.,%
\begin{equation}
\Pi ^{\mathrm{DimN}}(M^{2})=\int_{0}^{1}d\alpha \int_{0}^{1-a}d\beta \Pi ^{%
\mathrm{DimN}}(M^{2},\alpha ,\beta ),\ \ \Pi (M^{2})=\int_{0}^{1}d\alpha \Pi
^{\mathrm{DimN}}(M^{2},\alpha ).  \label{eq:A4}
\end{equation}%
In these expressions $\alpha $ and $\beta $ are Feynman parameters.

The perturbative and nonperturbative contributions of dimensions $3$, $4$, $%
5 $ and $7$ are terms of (\ref{eq:A2}) types. For relevant spectral
densities, we get%
\begin{equation}
\rho ^{\mathrm{pert.}}(s,\alpha ,\beta )=\frac{\Theta (L_{1})}{128\pi
^{6}L^{2}N_{1}^{7}}\left[ s\alpha \beta L-m_{b}^{2}N_{2}\right] ^{3}\left\{
3s\alpha \beta L^{2}+m_{b}^{2}N_{1}\left[ \alpha (\alpha -1)+\beta \left(
\beta -1\right) \right] \right\},
\end{equation}%
\begin{eqnarray}
&&\rho ^{\mathrm{Dim3}}(s,\alpha ,\beta )=\frac{m_{s}\left[ \langle \overline{s}s\rangle - 2\langle
\overline{u}u\rangle\right] }{8\pi
^{4}N_{1}^{5}}\Theta (L_{1})\alpha \beta \left\{ 2s^{2}\alpha \beta
L^{3}+m_{b}^{4}(\alpha +\beta )N_{1}^{2}-m_{b}^{2}s\left[ 2\beta
^{5}+2\alpha ^{2}(\alpha -1)^{3}\right. \right.  \notag \\
&&\left. \left. +\beta ^{4}(9\alpha -6)+\alpha \beta (\alpha -1)^{2}(9\alpha
-4)+2\beta ^{3}(3-11\alpha +8\alpha ^{2})-\beta ^{2}(2-17\alpha +31\alpha
^{2}-16\alpha ^{3})\right] \right\},
\end{eqnarray}%
\begin{eqnarray}
&&\rho ^{\mathrm{Dim4}}(s,\alpha ,\beta )=\frac{\langle \alpha _{s}G^{2}/\pi
\rangle }{768\pi ^{4}(1-\beta )L^{2}N_{1}^{5}}\Theta (L_{1})\alpha \left\{
6s^{2}\alpha \beta ^{2}(\beta -1)L^{3}\left[ 2\beta ^{2}+2(\alpha
-1)^{2}+\beta (5\alpha -4)\right] \right.  \notag \\
&&-m_{b}^{4}N_{1}^{2}(\alpha +\beta )\left[ 5\beta ^{4}+4\beta ^{3}(1-3\beta
)-8\alpha ^{3}(\alpha -1)-3\beta ^{2}(5-7\alpha +\alpha ^{2})+\beta
(6-9\alpha +3\alpha ^{2}-8\alpha ^{3})\right]  \notag \\
&&-sm_{b}^{2}\beta LN_{1}\left[ 4\beta (\beta -1)^{2}(3-6\beta +2\beta
^{2})+\alpha (\beta -1)(-12+78\beta -119\beta ^{2}+41\beta ^{3})\right.
\notag \\
&&\left. \left. +\alpha ^{2}(\beta -1)(36-119\beta +88\beta ^{2})+\alpha
^{3}(\beta -1)(49\beta -32)+16\alpha ^{4}(\beta -1)+8\alpha ^{5}\right]
\right\},
\end{eqnarray}%
\begin{equation}
\rho ^{\mathrm{Dim5}}(s,\alpha )=-\frac{m_{s}\left[ \langle \overline{s}%
g_{s}\sigma Gs\rangle -3\langle \overline{u}g_{s}\sigma Gu\rangle \right] }{%
48\pi ^{4}}\Theta (L_{2})(3m_{b}^{2}+s-4s\alpha +3s\alpha ^{2}),
\end{equation}%
\begin{equation}
\rho ^{\mathrm{Dim7}}(s,\alpha ,\beta )=-\frac{\langle \alpha _{s}G^{2}/\pi
\rangle m_{s}\left[ \langle \overline{s}s\rangle -2\langle \overline{u}%
u\rangle \right] }{96\pi ^{2}N_{1}^{3}}\alpha\beta L\Theta (L_{1}).
\end{equation}%
The dimension $6$ and $8$ terms have mixed compositions: they contain
components expressed through both $\rho ^{\mathrm{DimN}}(s)$ and $\Pi ^{%
\mathrm{DimN}}(M^2)$. For dimension $6$ term, we find%
\begin{equation*}
\Pi ^{\mathrm{Dim6}}(M^{2},s_{0})=\int_{\mathcal{M}%
^{2}}^{s_{0}}dse^{-s/M^{2}}\int_{0}^{1}d\alpha \rho ^{\mathrm{Dim6}%
}(s,\alpha )+\int_{0}^{1}d\alpha \int_{0}^{1-a}d\beta \Pi ^{\mathrm{Dim6}%
}(M^{2},\alpha ,\beta ),
\end{equation*}%
where%
\begin{equation}
\rho ^{\mathrm{Dim6}}(s,\alpha )=\frac{\Theta (L_{2})}{3\pi ^{2}}\left[
\langle \overline{s}s\rangle \langle \overline{u}u\rangle +\frac{g_{s}^{2}}{%
108\pi ^{2}}(\langle \overline{s}s\rangle ^{2}+\langle \overline{u}u\rangle
^{2})\right] (3m_{b}^{2}+s-4s\alpha +3s\alpha ^{2}),
\end{equation}%
\begin{eqnarray}
&&\Pi ^{\mathrm{Dim6}}(M^{2},\alpha ,\beta )=-\frac{\langle
g_{s}^{3}G^{3}\rangle m_{b}^{4}}{3840M^{2}\pi ^{6}\alpha ^{2}\beta
^{2}L^{4}N_{1}^{3}}\exp \left[ -\frac{m_{b}^{2}}{M^{2}}\frac{N_{1}(\alpha
+\beta )}{\alpha \beta L}\right]  \notag \\
&&\times \left\{ m_{b}^{2}(\alpha +\beta )N_{1}\left[ 5\beta ^{8}+2\beta
^{5}\alpha ^{2}(3-4\alpha )+2\beta ^{3}\alpha ^{4}(5-4\alpha )+3\beta \alpha
^{6}(\alpha -1)+5\alpha ^{6}(\alpha -1)^{2}+\beta ^{7}(3\alpha -10)\right.
\right.  \notag \\
&&\left. +\beta ^{4}\alpha ^{2}(-5+2(5-4\alpha )\alpha )-\beta ^{2}\alpha
^{4}(5+(\alpha -1)\alpha )-\beta ^{6}((3+\alpha )\alpha -5)\right]
+M^{2}\alpha \beta L  \notag \\
&&\times \left[ 14\beta ^{8}+14\alpha ^{6}(\alpha -1)+2\beta ^{3}\alpha
^{4}(4\alpha -3)+\beta ^{5}\alpha (\alpha -1)(8\alpha -17)+\beta ^{3}\alpha
^{4}(\alpha -1)(22\alpha -3)+\beta ^{7}(23\alpha -28)\right.  \notag \\
&&\left. \left. \beta \alpha ^{5}(\alpha -1)(23\alpha -17)+3\beta ^{4}\alpha
^{2}(1+2\alpha (\alpha -1))+2\beta ^{6}(7+\alpha (11\alpha -20))\right]
\right\}.
\end{eqnarray}%
Dimension $8$ contribution is given by expression%
\begin{eqnarray}
\Pi ^{\mathrm{Dim8}}(M^{2},s_{0}) &=&\int_{\mathcal{M}%
^{2}}^{s_{0}}dse^{-s/M^{2}}\int_{0}^{1}d\alpha \int_{0}^{1-a}d\beta \rho
_{1}^{\mathrm{Dim8}}(s,\alpha ,\beta )+\int_{\mathcal{M}%
^{2}}^{s_{0}}dse^{-s/M^{2}}\int_{0}^{1}d\alpha \rho _{2}^{\mathrm{Dim8}%
}(s,\alpha )  \notag \\
&&+\int_{0}^{1}d\alpha \int_{0}^{1-a}d\beta \Pi ^{\mathrm{Dim8}%
}(M^{2},\alpha ,\beta ).
\end{eqnarray}%
Here the relevant functions are equal to:%
\begin{eqnarray}
&&\rho _{1}^{\mathrm{Dim8}}(s,\alpha ,\beta )=\frac{\langle \alpha
_{s}G^{2}/\pi \rangle ^{2}}{1536\pi ^{2}N_{1}^{3}}\Theta (L_{1})\alpha \beta
(\alpha +\beta -1),\ \ \rho _{2}^{\mathrm{Dim8}}(s,\alpha )=-\frac{\langle
\overline{s}g_{s}\sigma Gs\rangle \langle \overline{u}u\rangle }{3\pi ^{2}}%
\Theta (L_{2})\ (1-4\alpha +3\alpha ^{2}),  \notag \\
&&\Pi ^{\mathrm{Dim8}}(M^{2},\alpha ,\beta )\ =\frac{\langle \alpha
_{s}G^{2}/\pi \rangle ^{2}m_{b}^{2}}{9216M^{4}\pi ^{2}\alpha ^{2}\beta
^{2}(1-\beta)L^{4}N_{1}^{3}}\exp \left[ -\frac{m_{b}^{2}}{M^{2}}\frac{%
N_{1}(\alpha +\beta )}{\alpha \beta L}\right]  \notag \\
&&\times \left\{ 96m_{b}^{4}\alpha ^{2}\beta ^{2}(\alpha +\beta )(\beta
-1)N_{1}^{2}\left[ 2\beta ^{2}+2\alpha (\alpha -1)+\beta (3\alpha -2)\right]
-M^{4}\alpha \beta L^{2}\left[ 4\beta ^{8}-\beta ^{7}(\alpha +16)\right.
\right.  \notag \\
&&+4\alpha ^{4}(\alpha -1)^{3}(2\alpha -1)+\beta ^{6}(24+4\alpha -206\alpha
^{2})+\beta \alpha ^{3}(\alpha -1)^{2}(-1-19\alpha +28\alpha ^{2})-\beta
^{5}(16+6\alpha -432\alpha ^{2}+23\alpha ^{3})  \notag \\
&&+\beta ^{4}(4+4\alpha -256\alpha ^{2}+67\alpha ^{3}-199\alpha ^{4})+\alpha
^{2}\beta ^{2}(-10+23\alpha -173\alpha ^{2}+313\alpha ^{3}-153\alpha ^{4})
\notag \\
&&\left. -\beta ^{3}\alpha (1-40\alpha +66\alpha ^{2}-385\alpha
^{3}+358\alpha ^{4})\right] +4m_{b}^{2}M^{2}(\alpha +\beta )LN_{1}\left[
\beta ^{4}(\beta -1)^{4}+2\alpha \beta ^{3}(\beta -1)^{3}(2\beta -1)\right.
\notag \\
&&-\alpha ^{2}\beta ^{2}(\beta -1)^{2}(-2+\beta (9+40\beta ))-\alpha
^{3}\beta (\beta -1)^{2}(-2+\beta (9+37\beta ))-\alpha ^{4}(\beta
-1)(1+\beta (-8+\beta (-31+85\beta )))  \notag \\
&&\left. \left. -\alpha ^{5}(\beta -1)(4\beta +1)(22\beta -3)+\alpha
^{6}(3-\beta (7+44\beta ))+\alpha ^{7}(\beta -1)\right] \right\}.
\end{eqnarray}%
The $\mathrm{Dim9}$ and $\mathrm{Dim10}$ contributions are exclusively of (%
\ref{eq:A4}) types. Thus, we have%
\begin{equation}
\Pi _{1}^{\mathrm{Dim9}}(M^{2},\alpha ,\beta )=-\frac{m_{b}^{2}\langle
g_{s}^{3}G^{3}\rangle m_{s}\left[ 2\langle \overline{u}u\rangle -\langle
\overline{s}s\rangle \right] }{2880M^{6}\pi ^{4}\alpha ^{4}\beta ^{4}(\beta
-1)L^{4}N_{1}^{2}}R_{1}(M^{2},\alpha ,\beta ),
\end{equation}%
and%
\begin{equation}
\Pi _{2}^{\mathrm{Dim9}}(M^{2},\alpha )=-\frac{\langle \alpha _{s}G^{2}/\pi
\rangle m_{s}\left[ \langle \overline{s}g_{s}\sigma Gs\rangle -3\langle
\overline{u}g_{s}\sigma Gu\rangle \right] }{3456M^{4}\pi ^{2}\alpha
^{4}(\alpha -1)^{2}}R_{2}(M^{2},\alpha ).
\end{equation}%
The dimension $10$ term has the following components:%
\begin{equation}
\Pi _{1}^{\mathrm{Dim10}}(M^{2},\alpha ,\beta )=\frac{\langle \alpha
_{s}G^{2}/\pi \rangle \langle g_{s}^{3}G^{3}\rangle m_{b}^{2}}{46080M^{6}\pi
^{4}\alpha ^{4}\beta ^{4}(\beta -1)L^{4}N_{1}^{2}}R_{1}(M^{2},\alpha ,\beta )
\end{equation}%
and%
\begin{equation}
\Pi _{2}^{\mathrm{Dim10}}(M^{2},\alpha )=\frac{\langle \alpha _{s}G^{2}/\pi
\rangle }{216M^{4}\alpha ^{4}(\alpha -1)^{2}}\left[ \langle \overline{s}%
s\rangle \langle \overline{u}u\rangle +\frac{g_{s}^{2}}{108\pi ^{2}}(\langle
\overline{s}s\rangle ^{2}+\langle \overline{u}u\rangle ^{2})\right]
R_{2}(M^{2},\alpha )
\end{equation}%
where functions $R_{1}(M^{2},\alpha ,\beta )$ and $R_{2}(M^{2},\alpha )$ are
given by formulas:
\begin{eqnarray}
&&R_{1}(M^{2},\alpha ,\beta )=\exp \left[ -\frac{m_{b}^{2}}{M^{2}}\frac{%
N_{1}(\alpha +\beta )}{\alpha \beta L}\right] \left\{ -6M^{4}\alpha
^{2}\beta ^{2}L^{3}\left[ \alpha ^{7}+2\alpha ^{6}(\beta -1)+\alpha
^{5}(\beta -1)^{2}+\beta ^{5}(\beta -1)^{2}\right] \right.  \notag \\
&&+m_{b}^{4}(\beta -1)N_{1}^{2}\left[ 5\beta ^{9}+5\alpha ^{7}(\alpha
-1)^{2}+\alpha ^{5}\beta ^{2}(\alpha -1)(5+2\alpha )+2\beta ^{8}(4\alpha
-5)+\beta \alpha ^{6}(\alpha -1)(8\alpha -5)\right.  \notag \\
&&+\beta ^{3}\alpha ^{4}(-5+(16-2\alpha )\alpha )+\beta ^{4}\alpha
^{3}(-5+4(5-4\alpha )\alpha )+\beta ^{5}\alpha ^{2}(-5-16(\alpha -1)\alpha
)+\beta ^{6}\alpha (5+3\alpha -9\alpha ^{2})  \notag \\
&&\left. +\beta ^{7}(5+\alpha (-13+2\alpha ))\right] +m_{b}^{2}M^{2}\alpha
\beta N_{1}^{2}\left[ 8\beta ^{8}+\alpha ^{5}(\alpha -1)^{2}(11\alpha
-8)+\beta ^{7}(23\alpha -24)+\beta ^{3}\alpha ^{4}(41\alpha -45)\right.
\notag \\
&&+3\beta ^{2}\alpha ^{4}(15-30\alpha +16\alpha ^{2})+\beta ^{6}(24-61\alpha
+30\alpha ^{2})+15\beta ^{4}\alpha (-1+2\alpha -\alpha ^{2}+\alpha ^{3})
\notag \\
&&\left. \left. +3\beta \alpha ^{4}(-5+19\alpha -25\alpha ^{2}+11\alpha
^{3})+\beta ^{5}(-8+53\alpha -60\alpha ^{2}+15\alpha ^{3})\right] \right\},
\end{eqnarray}%
and%
\begin{eqnarray}
R_{2}(M^{2},\alpha ) &=&\exp \left[ -\frac{m_{b}^{2}}{M^{2}\alpha (1-\alpha )%
}\right] \left[ -3M^{4}\alpha ^{3}(1+\alpha )(\alpha
-1)^{2}+8m_{b}^{4}(1-3\alpha +3\alpha ^{2})\right.  \notag \\
&&\left. -m_{b}^{2}M^{2}\alpha (16-51\alpha +48\alpha ^{2}+3\alpha ^{3})
\right].
\end{eqnarray}

In expressions above, $\Theta (z)$ is Unit step function. We have used also
the following short hand notations%
\begin{eqnarray}
N_{1} &=&\beta ^{2}+\beta (\alpha -1)+\alpha (\alpha -1),\ \ N_{2}=(\alpha
+\beta )N_{1},\ \ L=\alpha +\beta -1,\   \notag \\
L_{1} &=&\frac{(1-\beta )}{N_{1}^{2}}\left[ m_{b}^{2}N_{2}-s\alpha \beta L%
\right] ,\ \ \ \ L_{2}=s\alpha (1-\alpha )-m_{b}^{2}.
\end{eqnarray}

\end{widetext}

\end{document}